%% file: reeb-graph-arXiv.tex
\RequirePackage{tikz}

\documentclass[pdflatex,sn-mathphys]{sn-jnl}

\theoremstyle{thmstyleone}
\newtheorem{theorem}{Theorem}
\newtheorem{proposition}{Proposition}
\newtheorem{lemma}{Lemma}
\newtheorem{conjecture}{Conjecture}

\theoremstyle{thmstyletwo}%
\newtheorem{remark}{Remark}%

\theoremstyle{thmstylethree}%
\newtheorem{definition}{Definition}%

\usepackage{amsthm,amsmath,amssymb,amsfonts}

\usepackage{tikz-cd}

\newcommand {\mm}[1] {\ifmmode{#1}\else{\mbox{\(#1\)}}\fi}

\newcommand{\Mspace}       {\mm{\mathbb{M}}}
\newcommand{\Nspace}        {\mm{\mathbb{N}}}

\newcommand{\Rspace}        {\mm{\mathbb{R}}}

\newcommand{\updeg}        {\mm{d^{+}}}
\newcommand{\downdeg}        {\mm{d^{-}}}

\newcommand{\eps}        {\varepsilon}

\newcommand{\Scal}        {\mm{\mathcal{S}}}
\newcommand{\Tcal}        {\mm{\mathcal{T}}}
\newcommand{\Qcal}        {\mm{\mathcal{Q}}}

\newcommand{\M}        {\mm{U}}

\newcommand{\etal}{{et al.}}

\newcommand{\wrt}{{w.r.t.}}

\newcommand{\edit}[1]{{\textcolor{black}{#1}}}

\begin{document}

\title[Labeled Interleaving Distance for Reeb Graphs]
{\centering{Labeled Interleaving Distance\\
for Reeb Graphs}}

\author*[1]{\fnm{Fangfei} \sur{Lan}}\email{fangfei.lan@sci.utah.edu}

\author[2]{\fnm{Salman} \sur{Parsa}}
\email{s.parsa@depaul.edu}

\author[1]{\fnm{Bei} \sur{Wang}}\email{beiwang@sci.utah.edu}

\affil*[1]{\orgname{University of Utah}, \orgaddress{\city{Salt Lake City}, \state{Utah}, \country{USA}}}
\affil[2]{\orgname{DePaul University}, \orgaddress{\city{Chicago}, \state{Illinois}, \country{USA}}}

\abstract{
\input{sec-abstract}
}



\maketitle

\input{sec-introduction}

\input{sec-background}

\input{sec-labeled-interleaving}

\input{sec-algorithms}

\input{sec-geodesic}

\input{sec-conclusion}


\backmatter
\bmhead{Acknowledgments}
This work was partially supported by a grant from the Department of Energy (DOE) DE-SC0021015, and grants from the National Science Foundation (NSF) IIS-1910733 and IIS-2145499. 


\begin{appendices}
\input{sec-interleaving-details}

\end{appendices}


\input{reeb-graph-arXiv.bbl}



\end{document}

%% file: sec-abstract.tex
Merge trees, contour trees, and Reeb graphs are graph-based topological descriptors that capture topological changes of (sub)level sets of scalar fields.
Comparing scalar fields using their topological descriptors has many applications in topological data analysis and visualization of scientific data. 
Recently, Munch and Stefanou~introduced a labeled interleaving distance for comparing two labeled merge trees, which enjoys a number of theoretical and algorithmic properties. In particular, the labeled interleaving distance between merge trees can be computed in polynomial time. 
In this work, we define the labeled interleaving distance for labeled Reeb graphs. 
We then prove that the (ordinary) interleaving distance between Reeb graphs equals the minimum of the labeled interleaving distance over all labelings. We also provide an efficient algorithm for computing the labeled interleaving distance between two labeled contour trees (which are special types of Reeb graphs that arise from simply-connected domains). 
In the case of merge trees, the notion of the labeled interleaving distance was used by Gasparovic et al.~to prove that the (ordinary) interleaving distance on the set of (unlabeled) merge trees is intrinsic. 
As our final contribution, we present counterexamples showing that, on the contrary, the (ordinary) interleaving distance on (unlabeled) Reeb graphs (and contour trees) is not intrinsic. It turns out that, under mild conditions on the labelings, the labeled interleaving distance is a metric on isomorphism classes of Reeb graphs, analogous to the ordinary interleaving distance. This provides new metrics on large classes of Reeb graphs.

%% file: sec-introduction.tex
\section{Introduction}
\label{sec:introduction}

Topological descriptors such as merge trees, contour trees, and Reeb graphs capture topological changes of (sub)level sets of scalar fields.
Comparing scalar fields using their topological descriptors has a number of applications in topological data analysis (TDA) and visualization of scientific datasets such as combustion and molecular dynamics simulations, including symmetry detection~\cite{SaikiaSeidelWeinkauf2014,SaikiaSeidelWeinkauf2017,SridharamurthyMasoodKamakshidasan2020,ThomasNatarajan2011}, shape matching and retrieval~\cite{BarraBiasotti2013,BiasottiMariniMortara2003,HilagaShinagawaKohmura2001,SridharamurthyMasoodKamakshidasan2020,ZhangBajajBaker2004}; feature tracking~\cite{EdelsbrunnerHarerMascarenhas2008,SaikiaWeinkauf2017,SohnBajaj2006} and event detection~\cite{SaikiaSeidelWeinkauf2014,SridharamurthyMasoodKamakshidasan2020,LiPalandeYan2021}; clustering and classification~\cite{HilagaShinagawaKohmura2001,ZhangBajajBaker2004}, summarization~\cite{LohfinkWetzelsLukasczyk2020,YanWangMunch2020}, uncertainty visualization~\cite{GuntherSalmonTierny2014,WuZhang2013,YanWangMunch2020}, and interactive exploration~\cite{PocoDoraiswamyTalbert2015,YanWangMunch2020}; see~\cite{YanMasoodSridharamurthy2021} for a survey. 

\begin{figure}[!ht]
    \centering
    \includegraphics[width=0.5\linewidth]{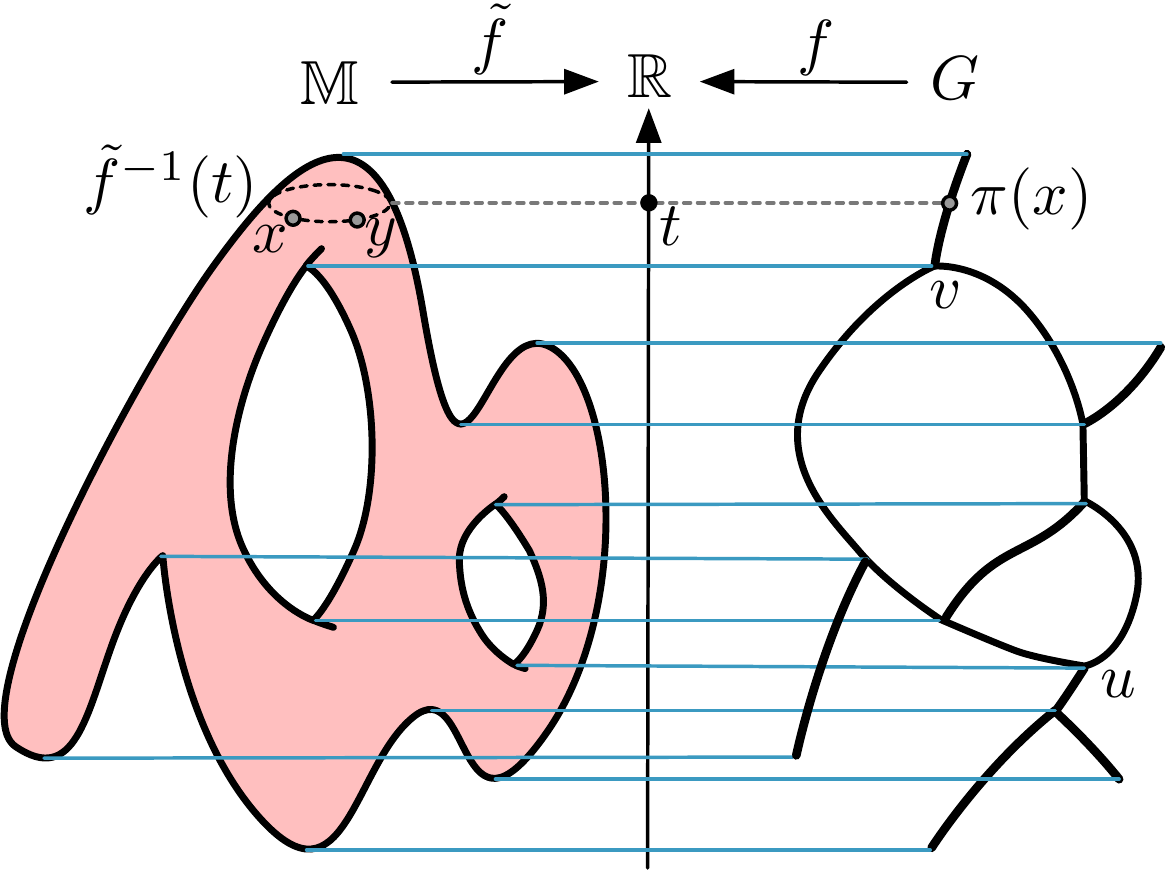}
    \caption{An example of a Reeb graph.}
    \label{fig:Reeb-graph}
\end{figure}

Given a continuous function $\tilde{f}: \Mspace \to \Rspace$ defined on a connected domain $\Mspace$, a Reeb graph records the connectivity of its \emph{level sets}. 
Two points $x, y \in \Mspace$ are considered \emph{equivalent} ({\wrt}~$\tilde{f}$), denoted as $x \sim y$, if $\tilde{f}(x) = \tilde{f}(y) = t$ and $x$ and $y$ belong to the same connected component of the level set $\tilde{f}^{-1}(t)$, for some $t \in \Rspace$. 
The \emph{Reeb graph} $\Mspace/{\sim}$ is the quotient space obtained by identifying equivalent points.
For well-behaved data, e.g.,~a Morse function on a manifold, $\Mspace/{\sim}$ is a graph $G$ that inherits a function $f$ from the original input data $(\Mspace, \tilde{f})$. See~\autoref{fig:Reeb-graph} for an example.    
A \emph{contour tree} is a special type of Reeb graph when the domain $\Mspace$ is simply connected. 
A merge tree, on the other hand, relies on equivalence relations among points in the \emph{sublevel sets} of $\tilde{f}$. That is, $x \sim y$, if $\tilde{f}(x) =\tilde{f}(y) = t$, and they belong to the same connected component of the sublevel set $\tilde{f}^{-1}(\infty, t]$, for some $t \in \Rspace$.

Since Reeb graphs generalize contour trees, and to some extent, merge trees, we mainly focus on comparative measures for Reeb graphs.  
A number of distances have been proposed for Reeb graphs and their variants, such as interleaving distance~\cite{BubenikDeSilvaScott2017,ChazalCohenSteinerGlisse2009,DeSilvaMunchPatel2016,MorozovBeketayevWeber2013,MunchStefanou2019,ChambersMunchOphelders2021}, functional distortion distance~\cite{BauerGeWang2014, BauerMunchWang2015}, functional contortion distance~\cite{BauerBjerkevikFluhr2022}, edit distance~\cite{DiFabioLandi2016,BauerFabioLandi2016, BauerLandiMemoli2020, SridharamurthyMasoodKamakshidasan2020},  Gromov-Hausdorff distance~\cite{CarriereOudot2017,TouliWang2019}, distances based on branch decompositions and matching~\cite{BeketayevYeliussizovMorozov2014,SaikiaSeidelWeinkauf2014}, and metrics for phylogenetic trees~\cite{CardonaMirRossello2013}.
The bottleneck distance~\cite{EdelsbrunnerHarer2010} is also defined for Reeb graphs~\cite{CarriereOudot2017} by computing the distance between their extended persistence diagrams~\cite{CohenSteinerEdelsbrunnerHarer2009}.
The stability of distances between Reeb graphs, their equivalence and the inequalities between them has been the topic of much research~\cite{DeSilvaMunchPatel2016,BauerGeWang2014,BauerLandiMemoli2020,BauerMunchWang2015,BakkeBjerkevik2021,BotnanLesnick2018}. Recently, it has been shown that the edit distance is universal, that is, it is stable and upper bounds all other stable distances~\cite{BauerLandiMemoli2020}, and hence is more discriminating.
There is now a vast literature on the subject of distances for Reeb graphs, see~\cite{BollenChambersLevine2021,YanMasoodSridharamurthy2021} for surveys.  

From the above distances, the bottleneck distance is polynomial-time computable.
The interleaving distance, the functional distortion distance and the edit distance are actually metrics on the isomorphism classes of Reeb graphs, hence most discriminatory. However, a major drawback of these metrics is that they are hard to compute.
Reeb graph $0$-interleaving is shown to be graph isomorphism complete~\cite{BjerkevikBotnan2018,DeSilvaMunchPatel2016}. 
And interleaving distances between multiparameter persistence modules are also NP-hard~\cite{BjerkevikBotnan2018,BjerkevikBotnanKerber2020}. 
As noted in~\cite{TouliWang2019}, the reduction of~\cite{AgarwalFoxNath2018} shows that it is NP-hard to approximate the interleaving distance for merge trees~\cite{MorozovBeketayevWeber2013} within a factor of 3. Therefore, it is also hard for Reeb graphs, since merge trees are special cases of Reeb graphs. 

Nevertheless, not all hope is lost.
\edit{Recently, Munch and Stefanou~\cite{MunchStefanou2019} showed that the $l^\infty$-cophenetic metric originally defined on phylogenetic trees is an example of an interleaving distance on labeled merge trees, where nodes of a merge tree are given labels from a fixed set.}
\edit{Gasparovic et al.~\cite{GasparovicMunchOudot2019} extended this work on the labeled interleaving distance of merge trees and proved that the (ordinary) interleaving distance is obtained as the minimum of the labeled distance over all possible labelings.}
The advantage of the labeled interleaving distance is that it can be computed efficiently in $O(n^2)$ ($n$ being the number of critical points of $f$). 
Such a notion makes the distance computation feasible in real-world  applications~\cite{YanMasoodRasheed2022,YanWangMunch2020}, and it appears to be a reasonable replacement for the (ordinary) interleaving distance, which is quite desirable. 
 
A labeled interleaving distance is applicable when there is a natural labeling for the nodes, or when a labeling may be inferred from the data. For instance, a climate simulation may give rise to an ensemble of scalar fields (e.g.,~temperature and pressure) simulated with different parameter settings. Each ensemble member is a scalar field defined on the same underlying mesh and gives rise to a slightly different merge tree. We may use the indices of mesh nodes as the labeling or infer correspondences between nodes of the merge tree based on similarities among features of their underlying scalar fields. 

Apart from computational efficiency, another desirable property for a distance between Reeb graphs is for the distance to be \emph{intrinsic}, i.e.,~the distance between any two Reeb graphs can be realized or approximated arbitrarily closely by a geodesic (a shortest continuous path) in the space of Reeb graphs~\cite{GasparovicMunchOudot2019, CarriereOudot2017}.  
An intrinsic distance is desirable for not only discrimination but also interpolation between a pair of Reeb graphs.   
Instead of studying the intrinsic-ness of a distance directly, Carriere and Oudot~\cite{CarriereOudot2017} studied the \emph{intrinsic metrics} induced by a number of distances between Reeb graphs, and showed that the intrinsic versions of Gromov-Hausdorff distance, interleaving distance, functional distortion distance and the bottleneck distance are all equivalent.  
To the best of our knowledge, it remains unknown whether most of the above distances proposed for Reeb graphs are intrinsic or not~\cite{CarriereOudot2017}. 
Gasparovic et al.~\cite{GasparovicMunchOudot2019} gave the first  positive answer in the setting of merge trees. They used the labeled interleaving distance to prove that the ordinary interleaving distance on merge trees is strictly intrinsic. This means that one can always find geodesics in the space of merge trees connecting two given merge trees, and such that their length approximates the distance arbitrarily closely. Therefore, there exists an average, \edit{albiet not necessarily unique}, of two merge trees, i.e., a merge tree that lies halfway between them in the sense of the interleaving distance.
Thus, for instance, in the climate simulation application, we may compute an average merge tree from an ensemble of merge trees. 
It appears hopeful that we could perform statistics on the space of contour trees and Reeb graphs, \edit{similar to} merge trees, using the interleaving distance.
 However, we show that the (ordinary, unlabeled) interleaving distance is not intrinsic. We arrived at this result by trying to prove the opposite, that is, to use the labeled interleaving distance to deduce, as in \cite{GasparovicMunchOudot2019}, that there are intermediary Reeb graphs. Hence the connection to the first part of our results.

\edit{Very recently, Bauer \etal~\cite{BauerBotnanFluhr2022} proved that the interleaving distance of sheaves on the real line is not intrinsic, which implies that the interleaving distance of Reeb graph is not intrinsic either. We were not
aware of this result. In contrast, our counterexamples are simple and we have a direct argument that proves that the interleaving distance of Reeb graphs is not intrinsic.}

We aim to generalize the results of~\cite{GasparovicMunchOudot2019} to the setting of Reeb graphs, succeeding partly:   
\begin{itemize} 
\item Our main contribution is the definition of a labeled interleaving distance for Reeb graphs (\autoref{sec:interleaving}) and~\autoref{theorem:essential-labeling}, which states that the ordinary interleaving distance between a pair of Reeb graphs can be obtained by taking the minimum of the labeled interleaving distance over all suitable labelings. Moreover, we show that under reasonable assumptions on the labelings, the resulting distance enjoys the same metric properties as the (unlabeled) interleaving distance (\autoref{t:exmetric} and \autoref{sec:metric-properties-details}).
\item We provide a simple $\tilde{O}(n^2)$ algorithm to compute the labeled interleaving distance between contour trees, where {$n$ is the number of nodes in the contour tree} (\autoref{sec:algorithms}). 
\item In the negative direction, we provide simple but not immediate  counterexamples and prove that the interleaving distance on Reeb graphs (and contour trees) is not intrinsic (\autoref{sec:counterexamples}). It follows that computing  averages of Reeb graphs is challenging at least with respect to the interleaving distance. 
\end{itemize}

%% file: sec-background.tex
\section{Background}
\label{sec:background}

\subsection{Reeb Graphs}
\label{sec:Reeb-graphs}

Assume we are given a well-behaved function $f: \Mspace \to \Rspace$, i.e., a Morse function on a manifold, a constructible \edit{$\Rspace$}-space~\cite{DeSilvaMunchPatel2016}, or a levelset-tame function~\cite{DeyWang2013} on a topological space. 
Following~\autoref{fig:Reeb-graph}, a \emph{Reeb graph} $\Mspace/{\sim}$ arises from a quotient map $\pi: (\Mspace, \tilde{f}) \to (\Mspace/{\sim}, f)$.
In our context, a Reeb graph is denoted as a pair $R=(G,f)$ where $G$ is a combinatorial finite multi-graph and $f: \vert G \vert \to \Rspace$ a function inherited from the input data defined as $f(\pi(x)) = \tilde{f}(x)$.  
Here, $G$ is a combinatorial object and $\vert G \vert$ is its underlying topological space; we sometimes use $G$ in place of $\vert G \vert$ for simplicity. 
In fact, our results apply to a more general setting, where a Reeb graph is defined to be a graph $G$ with function values associated with each node such that the function value on the edges is strictly monotone (with the maximum and minimum determined by the function values at the nodes)~\cite{DeSilvaMunchPatel2016}. For the rest of the paper, we work with this definition. Therefore, $f$ is strictly monotone on any edge of $G$. We also assume that $f$ is injective on the nodes.

The \emph{up-degree} $\updeg$ of a node $v$ of $G$ is the number of edges $uv$ such that $f(u)>f(v)$. 
The \emph{down-degree} $\downdeg$ is defined analogously. 
A node is called a \emph{split node} if its $\updeg > 1$ and $\downdeg = 1$. 
A node is called a \emph{join node} if $\updeg = 1$ and $\downdeg > 1$. 
In~\autoref{fig:Reeb-graph}, $u$ is a split node and $v$ is a join node. 
A node with $\updeg = \downdeg =1$ is a \emph{regular} node. 
A node with $\updeg = 0$ and $\downdeg =1$ is a \emph{(local) maximum}. 
A \emph{(local) minimum} is defined analogously.
A non-regular node is \emph{critical} and corresponds to a critical point of $\tilde{f}$ (i.e., for a smooth $\tilde{f}$).  

In degenerate scenarios, a Reeb graph might contain a degenerate node, for instance, with $\updeg = 0$ and $\downdeg = 2$. For technical reasons and for simplifying our proofs, we consider such a node a superposition of two nodes, one with $\updeg = 1$ and $\downdeg = 2$, and another with $\updeg = 0$ and $\downdeg = 1$. 
We apply similar operations for nodes with $\updeg \geq 2$ and $\downdeg \geq 2$.   
This way, a single degenerate node of the Reeb graph consists of several superposed simpler nodes.
We consider the nodes in superposition to be connected by edges of length $0$.
 
When it is clear from the context, we abuse the notations and denote by $R$ some geometric realization of the Reeb graph $R=(G,f)$ (e.g.,~embedded in $\Rspace^3$ for visualization) such that $f$ is shown as a height function in the vertical direction; see~\autoref{fig:Reeb-graph}. In this way, we could talk about an \emph{arc} being a subset of $R$ that does not necessarily start or end at a node. In addition, we denote a geometric realization of the graph $G$ also by $G$.

\subsection{Interleaving Distance Between Reeb Graphs}
\label{sec:ordinary-interleaving}

We recall the (ordinary) interleaving distance in a geometric way following notations in~\cite{ChambersMunchOphelders2021, DeSilvaMunchPatel2016}. 
We refrain from using categorical language as long as it is not needed in our arguments; see~\cite{DeSilvaMunchPatel2016} for a categorical interpretation of Reeb graphs. 

\begin{figure}[!ht]
    \centering
    \includegraphics[width=0.55\linewidth]{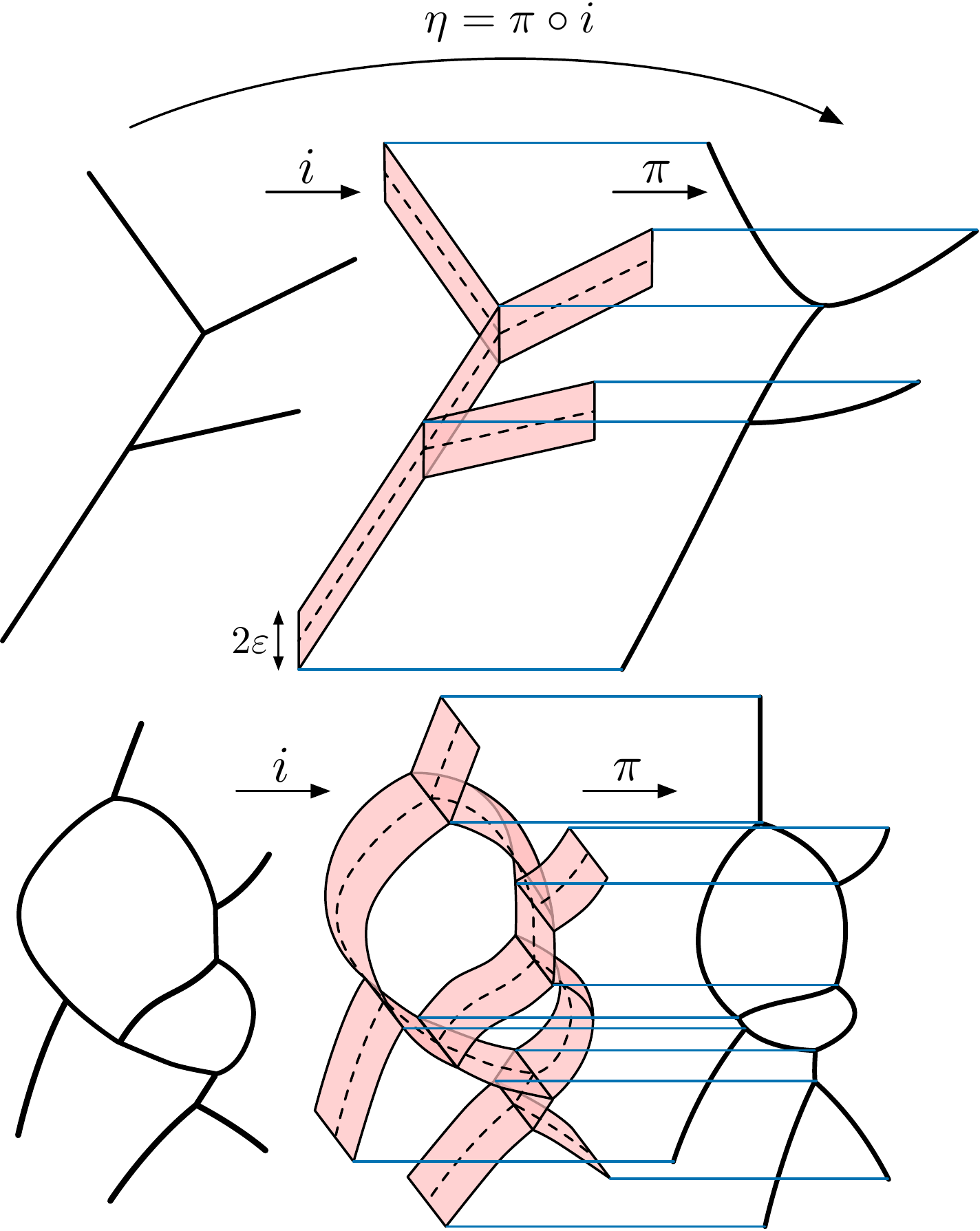}
    \caption{Examples for a contour tree (top) and a Reeb graph (bottom). From left to right: a Reeb graph $R=(G,f)$, its $\eps$-thickening $G \times [-\eps, \eps]$ with a function $F^{\eps}$, and the Reeb graph $\mathcal{S}^{\eps}(R)$ of the $\eps$-thickening.}
    \label{fig:smoothing}
\end{figure}

Let $R_1=(G_1, f_1)$ and $R_2=(G_2, f_2)$ be two Reeb graphs.
A \emph{morphism} of Reeb graphs, $h: R_1 \to  R_2$, is a continuous function $h:G_1 \to G_2$ that is \emph{function-preserving},  i.e., $f_2(h(x))=f_1(x)$.

For a Reeb graph $R=(G,f)$ and $\eps \geq 0$, we define the \emph{$\eps$-thickening} of $R$,  denoted $\Tcal^\eps(R)$, to be $G \times [-\eps, \eps]$  with the product topology, together with the function $F^{\eps}: G \times [-\eps, \eps] \to \Rspace$, $F^{\eps}(x,t) \mapsto f(x) + t$. 
We then define the \emph{$\eps$-smoothing} $\Scal^\eps(R)$ to be the Reeb graph of $\Tcal^\eps(R)$ with respect to $F^{\eps}$. 
$\Scal^\eps(R)$ is equipped with a function $f^{\eps}$ that is inherited from $F^{\eps}$.
See~\autoref{fig:smoothing} for examples involving a contour tree (top) and a Reeb graph (bottom) respectively. The $\eps$-thickening of the Reeb graph in~\autoref{fig:smoothing} (bottom) is tilted slightly to reveal its structure. 
For simplicity, we abbreviate $R^\eps = \Scal^\eps(R)$.

By construction, there is an inclusion $i: G \to G \times [-\eps, \eps]$, $x \mapsto (x,0)$, as well as a quotient map $\pi: G \times [-\eps, \eps] \to R^\eps$; their composition gives rise to an \emph{$\eps$-shift} morphism, $\eta:=\pi \circ i : R \to R^\eps$, which assigns to the point $x \in G$ the connected component of $(x,0) \in G \times [-\eps, \eps]$ in the pre-image of $F^\eps$, see \cite{DeSilvaMunchPatel2016} for details.

If $\phi: R_1 \to R_2^\eps$ is a morphism, then, for every $\delta\geq 0$ including $\delta=\eps$, there exists a morphism $\phi^\delta: R_1^\delta \to R_2^{\delta+\eps}$.\footnote{This map is defined first as a function-preserving map from the thickening $T^\delta(R_1)$ into the thickening $T^{\delta}(R^\eps_2)$ as $\Phi^\delta(x,t) = (\phi,t)$. This map will induce the map $\phi^{\delta}$ on the Reeb graphs.} 
Analogously, if $\psi: R_2 \to R_1^\eps$ is a morphism, then there exists a morphism $\psi^\eps: R_2^\eps \to R_1^{2\eps}$.  
Similarly, given a morphism $\eta_1: R_1 \to R_1^{\eps}$, there exists a morphism $\eta_1^{\eps}: R_1^{\eps} \to R_1^{2\eps}$; $\eta_2^{\eps}$ is defined analogously. 

\begin{definition}
\label{definition:interleaving}
Let $R_1$ and $R_2$ be a pair of Reeb graphs. 
An $\eps$-interleaving between $R_1$ and $R_2$ is given by two morphisms, 
$
\phi: R_1 \to R_2^\eps, \quad \psi: R_2 \to R_1^\eps,    
$
such that the following diagram commutes.
\begin{equation}
  \begin{tikzcd}
  R_1
    \ar[r,"\eta_1"]
    \ar[dr, "\phi", very near start, outer sep = -2pt]
  & R_1^\eps
    \ar[r, "\eta_1^\eps"]
    \ar[dr, "\phi^{\eps}", very near start, outer sep = -2pt]
  & R_1^{2\eps}
  \\
  R_2
    \ar[r, swap,"\eta_2"]
    \ar[ur, crossing over, "\psi"', very near start, outer sep = -2pt]
  & R_2^\eps
    \ar[r, swap,"\eta_2^\eps"]
    \ar[ur, crossing over, "\psi^{\eps}"', very near start, outer sep = -2pt]
  & R_2^{2\eps}
  \end{tikzcd}
\end{equation}

\noindent Equivalently, see~\cite[Definition 4.38]{DeSilvaMunchPatel2016}, we require that the following two relations hold,
\begin{equation}
\label{eq:tworelations}
    \begin{split}
        \phi^\eps \circ \psi &= \eta_2^{2\eps}\\
        \psi^\eps \circ \phi &= \eta_1^{2\eps},
    \end{split}
\end{equation}
where $\eta_1^{2\eps} = \eta_1^\eps \circ \eta_1: R_1 \rightarrow R_1^{2\eps}$ and $\eta_2^{2\eps} = \eta_2^\eps \circ \eta_2: R_2 \rightarrow R_2^{2\eps}$ are the $2\eps$-shift morphisms.

The \emph{interleaving distance} is defined to be 
\begin{equation}
d_I(R_1, R_2) := \inf \{\eps\geq 0 \mid \text{there exists an $\eps$-interleaving between } R_1\; \text{and}\; R_2\}.
\end{equation}
\end{definition}

Smoothing by $2\eps$ is \edit{isomorphic to} smoothing twice by $\eps$.
It is known that if we consider $\eps$-smoothings of a Reeb graph $R$ for increasing values of $\eps \geq 0$, the minima and join nodes move downwards (i.e.,~they move to points with lower function values in the smoothed Reeb graph), and maxima and split nodes move upwards, see~\cite{BauerMunchWang2015, AlharbiChambersMunch2021} for a proof. 
Recall that we consider a degenerate node (e.g. with $\updeg>1$ and $\downdeg>1$) as a superposition of a join node and a split node. When smoothed by $\eps$ these nodes give rise to two nodes, one split and the other join, that are separated by $2\eps$. 

If $T:=R$ is a contour tree, \edit{every} node $v$ of $T$ \edit{corresponds} to a node in $T^\eps$ which lies $\eps$ above or below in function value. We denote this correspondence by $s: V(T) \rightarrow V(T^\eps)$.
For nodes in a superposition, each superposed node has its image under $s$. 
This is the reason behind the idea of superposition, since each superposed node moves differently when smoothed. 

If $R$ is a Reeb graph, a split node $v$ and a join node $w$ might cancel each other in the smoothed graph, and a loop might   disappear. 
In this case, there are no node correspondences in the smoothed  graph. 
For a split vertex $v$, we define $s(v)=\pi((v,\eps)) \in R^\eps$, and, for a merge vertex $w$, we define $s(w)=\pi((w, -\eps)) \in R^\eps$. In this way, $s:V(R) \rightarrow R^{\eps}$ is defined on all nodes. 



\subsection{Intrinsic Metrics and Geodesic Spaces}
Two Reeb graphs $R_1$ and $R_2$ are \emph{isomorphic} if 
there are function-preserving continuous maps $h: R_1 \xrightarrow{} R_2$ and $g: R_2 \xrightarrow{} R_1$ such that $f \circ g$ and $g \circ f$ are identity maps. 
This is equivalent to $d_I(R_1,R_2)=0$ since by definition, $h$ and $g$ define a 0-interleaving.

The interleaving distance is a metric if we identify isomorphic Reeb graphs, see \cite[Proposition~4.8]{DeSilvaMunchPatel2016}. 
Given any metric $d$ over the set of (isomorphism classes) of Reeb graphs, we obtain the \textit{space of Reeb graphs}. Given a continuous path $\gamma$ (i.e.,~continuous {\wrt}~$d$) into the space of Reeb graphs, where $\gamma(0)=R_1$ and $\gamma(1)=R_2$, the length $\ell_d(\gamma)$ induced by $d$  is defined as
\[
    \ell_d(\gamma) = \sup_{n,\Gamma} \sum_{i=1}^{n} d(\gamma(\Gamma(i)), \gamma(\Gamma(i+1)),  
\]
where $n$ ranges over all $\Nspace$ (natural numbers) and $\Gamma$ ranges over all partitions of $[0,1]$, i.e.~all ordered sets of $n$ points $\Gamma = (\Gamma(1), \ldots, \Gamma(n))$ such that $\Gamma(i) \in [0,1]$; see~\cite{BuragoBuragoIvanov2001, GasparovicMunchOudot2019}. 

Given a metric $d$, the \emph{intrinsic metric} $\hat{d}$ induced by $d$ is defined as 
$\hat{d} (R_1, R_2) = \inf_{\gamma} \ell_d(\gamma);$
it is the infimum of the lengths of all paths from one point to another.
It is known that $d \leq \hat{d}$. 
A metric $d$ is \emph{intrinsic} if it is always equal to its intrinsic version $\hat{d}$; in this case, the metric space is said to be a \emph{length space}. 
A metric space is a \emph{geodesic space} if any two points in the space can be connected by a curve of length equal to the distance between the two points; the metric is then called \emph{strictly intrinsic}~\cite{ BuragoBuragoIvanov2001,CarriereOudot2017,GasparovicMunchOudot2019}. 
A strictly intrinsic metric is of course intrinsic, thus a geodesic space is necessarily a length space.

%% file: sec-labeled-interleaving.tex
\section{Labeled Interleaving Distance Between Reeb Graphs}
\label{sec:interleaving}

In this section, we first define the labeled interleaving distance between a pair of labeled Reeb graphs and prove its properties. 
Then we show that such a distance generalizes the labeled interleaving distance between merge trees. 

\subsection{Labeled Reeb Graphs and Their Distance}

\begin{definition}
\label{definition:labeled-Reeb}
Let $L$ be a finite set of \emph{labels}. 
For simplicity, $L = [N]:=\{1,\dots,N\}$.
Let $V=V(R)$ denote the set of nodes of a Reeb graph $R = (G,f)$. A \emph{labeling} of $R$ is a function $\lambda : L \to V$. We call the triple $R_\lambda = (G, f, \lambda)$ an \emph{$L$-labeled Reeb graph}, or simply a \emph{labeled Reeb graph} when the set of labels $L$ is clear from context. $R_\lambda$ is \emph{fully-labeled} if $\lambda$ is surjective. Note that $\lambda$ is not necessarily injective. 
A \emph{morphism} between labeled Reeb graphs is defined to be the morphism of the underlying unlabeled Reeb graphs. 
\end{definition}

\begin{figure}[!ht]
    \centering
    \includegraphics[width=0.5\linewidth]{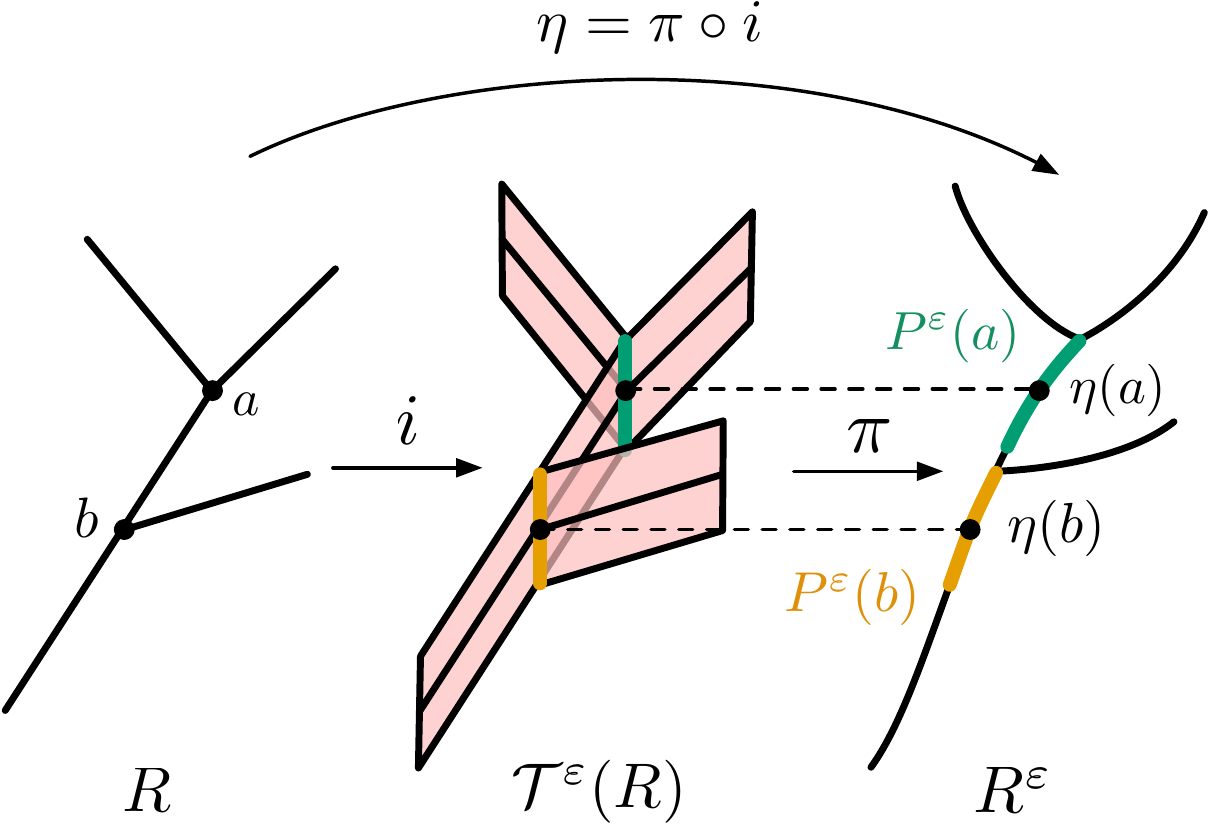}
    \caption{The $\eps$-path-neighborhoods of node $a$ and $b$ are highlighted in green and yellow in the right-most figure, respectively.}
    \label{fig:path-neighborhood}
\end{figure}

\begin{definition}
\label{definition:path-nbhd}
Let $v \in V(R)$ be a node in the Reeb graph $R = (G,f)$. The \emph{$\eps$-path-neighborhood} of $v$, denoted $P^\eps(v)$, is $\pi(\{v\} \times [-\eps, \eps] ) \subset R^\eps$ in the $\eps$-smoothed Reeb graph $R^\eps$. Here $\pi: G\times [-\eps, \eps] \to R^\eps$ is the quotient map. For any $x\in R$, $P^\eps(x)$ is defined similarly.
\end{definition}

As illustrated in~\autoref{fig:path-neighborhood}, we observe that for any point $x \in R$ and any $\eps \geq 0$, $P^\eps(x)$ is a monotonic path in $R^\eps$ such that $f^\eps(P^\eps(x)) = [f(x)-\eps, f(x)+\eps)]$. 
In other words, $P^\eps(x)$ is a monotonic path centered on the image of $x$ in $R^\eps$, i.e., centered on $\eta(x):=\pi \circ i(x)$.

\begin{definition}
\label{definition:labeled-interleaving}
Let $R_{1,\lambda_1}=(G_1, f_1,\lambda_1)$ and $ R_{2,\lambda_2} = (G_2,f_2, \lambda_2)$ be two $L$-labeled Reeb graphs, and let $\eps\geq 0$. We say a pair of morphisms $\phi: R_1 \to R_2^\eps$ and $\psi: R_2 \to R_1^\eps$ define a \emph{labeled $\eps$-interleaving} between $R_{1,\lambda_1}$ and $R_{2,\lambda_2}$ if the following hold:
\begin{enumerate}
    \item $\phi$ and $\psi$ define an $\eps$-interleaving between $R_1$ and $R_2$.
    \item For each $\ell \in L$, we have the following label-preserving properties, 
\begin{equation}
\label{eq:label-preserving}
        \begin{split}
            \phi^\eps(s(\lambda_1( \ell))) &\in P^{\eps}(s(\lambda_2(\ell))),\\
 \psi^\eps(s(\lambda_2(\ell))) &\in P^{\eps}(s(\lambda_1(\ell))).
        \end{split}
    \end{equation} 
\end{enumerate} 
\noindent In the the above formulae, $P^\eps(s(\lambda_1(\ell))) \subset R_1^{2\eps}$ and $P^\eps (s(\lambda_2(\ell))) \subset R_2^{2\eps}.$ 
\noindent The \emph{labeled interleaving distance} between $R_{1,\lambda_1}$ and $R_{2,\lambda_2}$ is defined as
\begin{align}
& d^L_I(R_{1,\lambda_1}, R_{2,\lambda_2}) \nonumber\\
& = \inf \{ \eps\geq 0 \mid \text{there exists a labeled $\eps$-interleaving between } R_{1,\lambda_1}\; \text{and}\; R_{2,\lambda_2}\;\}.
\end{align}
\end{definition} 
Recall that the function $s$ maps the nodes of a Reeb graph to the points of the $\eps$-smoothed Reeb graph (see the end of~\autoref{sec:interleaving}).
Intuitively, it outputs the point where the original node would arrive at after moving up or down by $\eps$.
For simplified notation, we refer to all such functions by $s$, instead of naming them differently based on their domain or co-domain. 

From~\autoref{definition:labeled-interleaving}, it follows easily that for all $R_1, R_2, \lambda_1$, and $\lambda_2$, we have, 
$$d^L_I(R_{1,\lambda_1}, R_{2,\lambda_2}) \geq d_I(R_1, R_2).$$
If the label set $L$ is empty, then the second condition in~\autoref{definition:labeled-interleaving} is vacuous and the labeled interleaving distance equals the unlabeled one. 

\begin{figure}[!ht]
    \centering
    \includegraphics[width=0.65\linewidth]{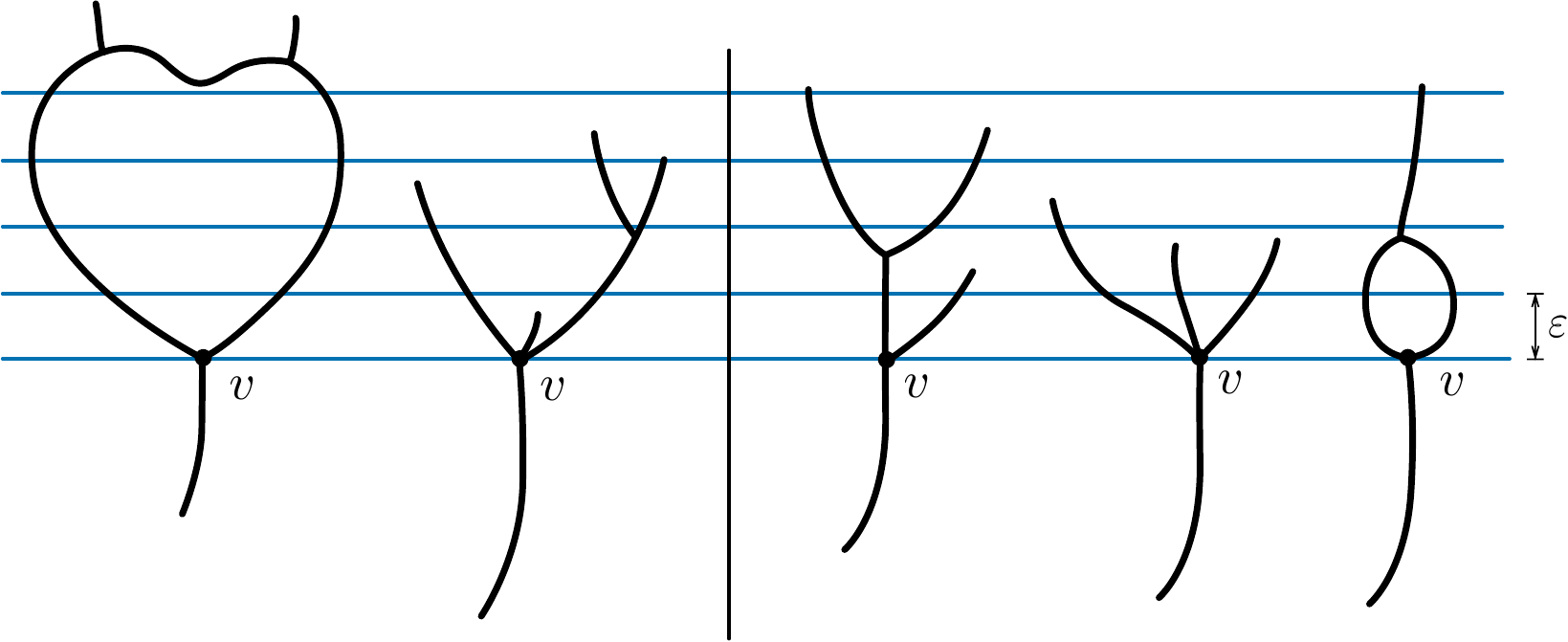}
    \caption{$\eps$-essential (left) and $\eps$-inessential nodes (right).}
    \label{fig:essential}
\end{figure}

Let $R_1$ and $R_2$ be two Reeb graphs. It turns out that the existence of maps defining an $\eps$-interleaving between $R_1$ and $R_2$ does not depend on all of the nodes and edges of $R_1$ and $R_2$. Intuitively, it depends only on the topological features that are significant enough {\wrt} to $\eps$. We make this precise by defining the notion of $\eps$-essential and $\eps$-inessential nodes.

\begin{definition}
\label{definition:essential}
Let $R=(G,f)$ be a Reeb graph and let $\eps>0$ be given. A split node $v$ is called \emph{$\eps$-inessential} if one of the following holds: 
\begin{enumerate}
\item There exists a join node $w$ and two paths $Q_1$ and $Q_2$ such that each $Q_i$ $(i \in \{1, 2\})$ joins $v$ to $w$ with $f(v) \leq f(x) \leq f(v)+4\eps$, for all $x\in Q_i$.
\item Let $\Qcal$ be the set of all paths $Q=vu$ \edit{such that $f(u)\geq f(v)+2\eps$ and $\forall x \in Q, f(x)\geq f(v)$}. Then all paths in $\Qcal$ use the same
    outgoing edge from $v$.
\end{enumerate}
An  $\eps$-inessential join node is defined analogously. 
Then \emph{$\eps$-essential} nodes are all maxima, minima, join and split nodes that are not $\eps$-inessential. 
\end{definition}

The first condition of the $\eps$-inessential definition implies that there is a loop of height at most $4\eps$ such that $v$ is the lowest point of the loop. The second condition indicates that the parts of the graph reachable from $v$ via paths that do not go below $f(v)$ either have heights at most $2\eps$, or they are reached using the same edge incident on $v$; see \autoref{fig:essential}.

We call a labeling an \emph{$\eps$-essential labeling} if every $\eps$-essential node is labeled. 
The following main theorem shows that we obtain the (ordinary) interleaving distance between two Reeb graphs if we consider all possible $\eps$-essential labelings. 
\begin{theorem}
\label{theorem:essential-labeling}
Let $R_1$ and $R_2$ be two Reeb graphs and set $\eps = d_I(R_1,R_2)$. There exist a set $L$ of labels and $\eps$-essential labelings $\lambda_1$ and $\lambda_2$ such that for $L$-labeled Reeb graphs $R_{1,\lambda_1}$ and $R_{2,\lambda_2}$, we have 
$$d_I(R_1, R_2) = d^L_I(R_{1,\lambda_1}, R_{2,\lambda_2}).$$ 
\end{theorem}
The reason for considering $\epsilon$-essential labelings in \autoref{theorem:essential-labeling} is to present a label set $L$ and labelings that cover a good portion of the nodes; the theorem
would be trivial for $L=\emptyset$.

\begin{figure}[!ht]
    \centering
    \includegraphics[width=0.7\linewidth]{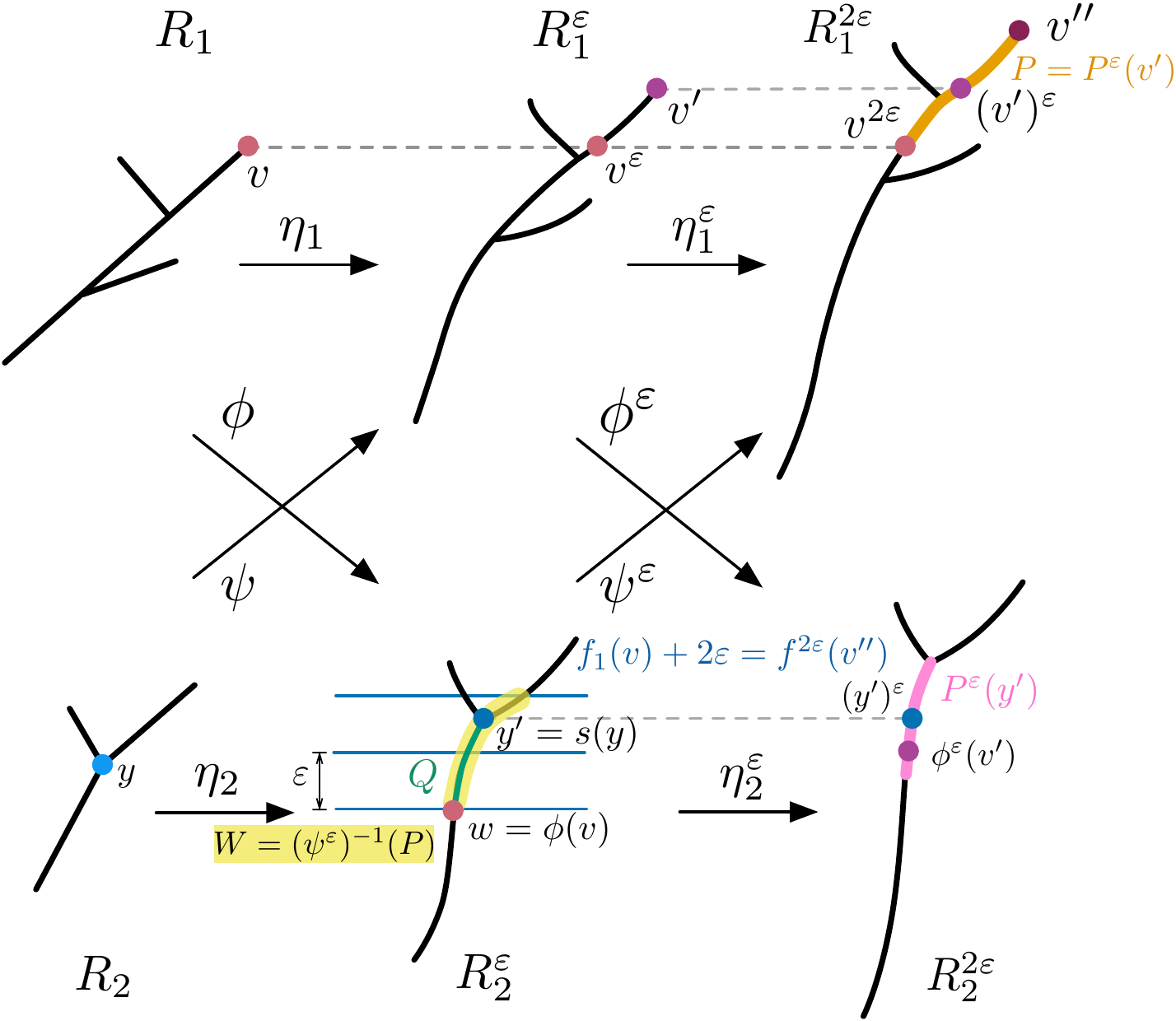}
    \caption{An illustration for the proof of~\autoref{theorem:essential-labeling}.}
    \label{fig:theorem1}
\end{figure}

We first sketch the proof and then present it in full detail.
Let $\eps = d_I(R_1,R_2)$. We consider the case of a maximum or an essential split node $v$ in $R_1$, as illustrated in \autoref{fig:theorem1}. Then $v'' = s(s(v))$ is also a maximum or split node in $R_1^{2\eps}$. Let $P = P^\eps(s(v)) \subset R_1^{2\eps}$ be the $\eps$-path neighborhood of $s(v)$. We then consider the space $W = (\psi^\eps)^{-1}(P) \subset R_2^\eps$. From the properties of $\eps$-interleaving, we deduce that $w=\phi(v) \in W$. Next, we walk upward in $W$: starting from $w$ until reaching a split node or a maximum. We argue that one of these nodes (i.e.,~a split node or a maximum) is reachable inside $W$. We denote such a maximum or split node by $s(y)$ for some $y \in R_2$, and pair $v$ with $y$. These pairs provide the desired labeling.

\begin{proof}
Let $\eps = d_I(R_1, R_2)$ be realized by morphisms $\phi: R_1 \to R^\eps_2$, $\psi: R_2 \to R_1^\eps$. Let $V_1 = V(R_1)$ and $V_2 = V(R_2)$ be the node sets. We will construct a subset $B \subset V_1 \times V_2$ and use $B$ to define a labeling on $R_1$ and $R_2$.

As illustrated in~\autoref{fig:theorem1}, let $v \in V_1$ be either a maximum or an essential split node. Then there is also a maximum or a split node $v''=s(s(v))$ in $R_1^{2\eps}$ such that $f^{2\eps}_1(v'')= f_1(v)+2\eps.$
Let $w=\phi(v)$ in $R_2^\eps$. Also, let $v'=s(v)$ and let $P=P^{\eps}(v')$ be the $\eps$-path neighborhood of $v'$ in $R^{2\eps}_1$. $P$ is a path with endpoints $v''$ from above and $v^{2\eps}$ from below. For simplicity, we denote the image $\eta_1(v_i)$ for any node $v_i \in R_1$ by $v_i^\eps$. Then, $\eta_1^{\eps}(\eta_1(v)) = v^{2\eps}$. Let $W \subset R^\eps_2$ be the pre-image $W= (\psi^\eps)^{-1}(P)$. We have $w\in W$, since $\psi^\eps(w)=\psi^\eps(\phi(v))=\eta_1^{\eps}(\eta_1(v))$. 

We walk up (i.e. in the direction of increasing function values) from $w \in W$ until we reach
a split node or a maximum. We claim that one of these two possibilities must occur while the function value is at most $f_1(v)+2\eps=f_1^{2\eps}(v'')$. If $v$ is a maximum then the point with the maximum value in $W$ must be a maximum node with function value at most $f_1^{2\eps}(v'')$. It follows that in this case we either reach a split mode or a maximum. 
If $v$ is an essential split node, we claim that for a sufficiently small $\delta>0$, there are two points $a$ and $b$ in $R_1$ with $f_1(a)=f_1(b)=f_1(v)+2\eps+\delta$
such that $a^{2\eps} \neq b^{2\eps}$. To see this, we argue as follows. 
We move up along the two different branches of the split node $v$. If the branches join before advancing by value $2\eps+\delta$, then this satisfies the first condition of inessentiality (\autoref{definition:essential}), violating our assumption that $v$ is an essential split node. 
Thus, $a\neq b$ at function value $f(v)+2\eps+\delta$.
Since the split node moves up at most $2\eps$ in $R_1^{2\eps}$, $a^{2\eps} = b^{2\eps}$ is only possible if there exists a join node $j \in R_1$ that moves past the value of $a$ and $b$ from a higher function value. Since $j$ moves down by $2\eps$, $j''$ will merge with the split node $v''$ if it moves at most $\delta$ more. Thus, there must be a loop with height at most $4\eps + \delta$. Since $\delta$ is small, the loop has height at most $4\eps$, which again satisfies the first condition of inessentiality. We reach a contradiction.
Therefore, we have $a^{2\eps}\neq b^{2\eps}$.

Since $\psi^\eps \circ \phi = \eta_1^{\eps} \circ \eta_1$ and $a^{2\eps}\neq b^{2\eps}$, then $\phi(a) \neq \phi(b)$. 
This implies that we will see a split node when moving up from $w$ before passing the function value $f_1(v)+2\eps$ (in fact, before passing $f_1(v)+2\eps+\delta$ by the above argument for all small $\delta>0$, 
which implies before passing $f_1(v)+2\eps$). It might be that the split node has value $f_1(v)+2\eps$. This proves our claim for an essential split node.

Regardless of the type of the node $v$, let $y'$ be the split node or the maximum we reach when moving up from $w$. Note that it could be that $y'=w$.
Then $f^\eps_2(y')\leq f^{2\eps}(v'') =f_1(v)+2\eps$, meaning that we travel by at most $2\eps$ in function values. $y'$ corresponds to a split or a maximum $y$ in $R_2$ such that $y'=s(y)$ (recall our convention on superposition of nodes). We add $(v,y)$ to $B$, indicating that $v\in R_1$ and $y \in R_2$ are assigned the same label. We have $\psi^\eps(y') \in P^\eps(v')$ by construction. 
If $y$ is an inessential split mode, we have completed the proof. Otherwise, we need to show that $\phi^\eps(v') \in P^\eps(y')$ to stop considering the node $y$.

Let $Q$ be the path that we have traversed connecting $w$ to $y'$. Since $Q$ is a monotone path with height at most $2\eps$, applying~\autoref{lemma:split-nbhd} below with $v$ and $z=y'$ implies that $\phi^\eps(v') \in P^\eps(y')$. 

We analogously pair all minima and essential join nodes of $R_1$ with minima and join nodes of $R_2$.
After pairing all the essential nodes of $R_1$ with nodes of $R_2$, we repeat the above process for the essential nodes of $R_2$ that are not paired yet. In the end, we obtain a set $B$ containing all paired nodes. We choose $B$ to be our label set $L$.
By construction the morphisms $\phi$ and $\psi$ define a labeled $\eps$-interleaving {\wrt} this choice of labels.
\end{proof}

\begin{figure}[!ht]
    \centering
    \includegraphics[width=0.8\linewidth]{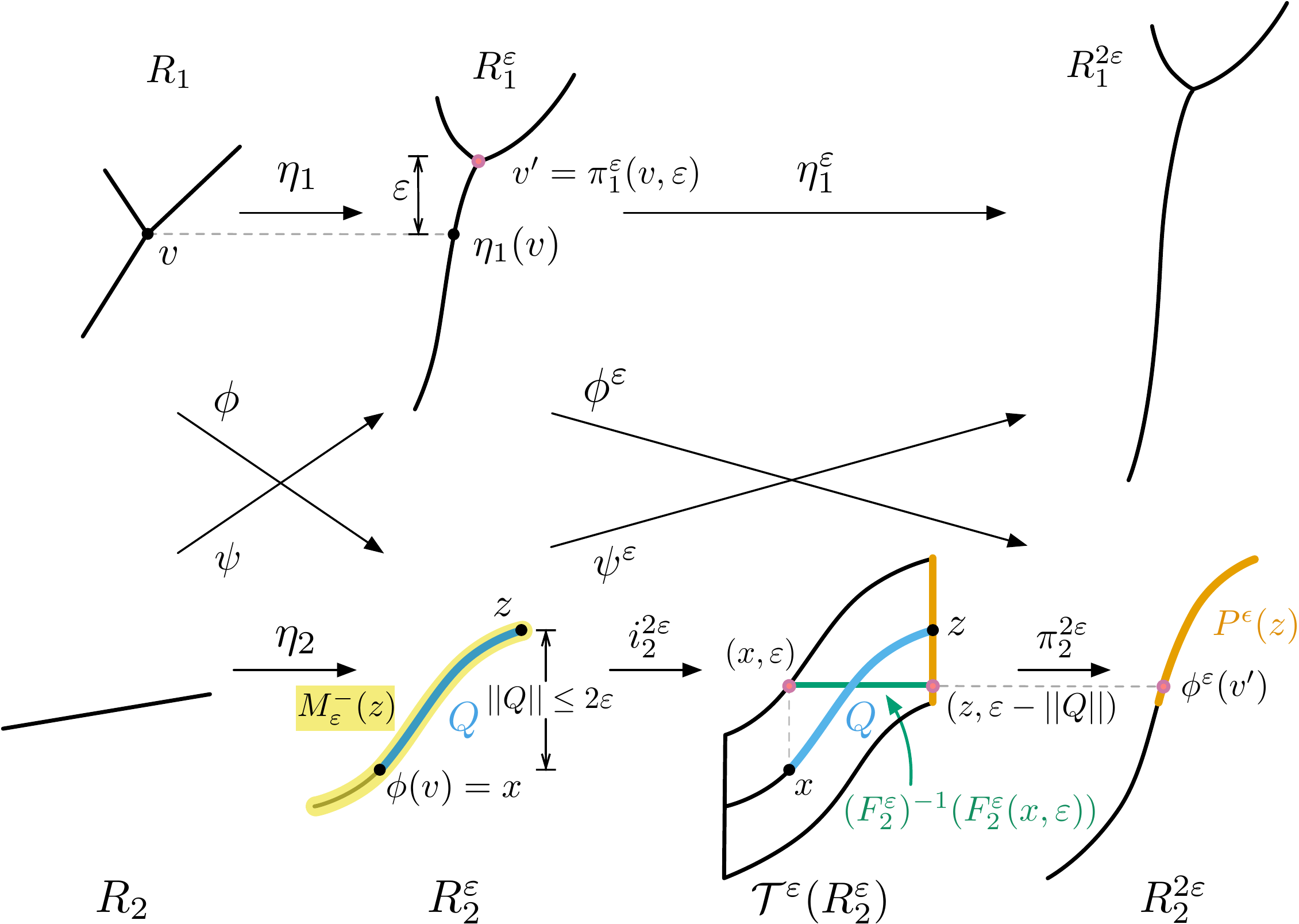}
    \caption{An illustration for the proof of~\autoref{lemma:split-nbhd}.}
    \label{fig:lemma1}
\end{figure}

\begin{lemma}
\label{lemma:split-nbhd}
Let $\phi$ and $\psi$ define an $\eps$-interleaving between $R_1$ and $R_2$.
Let $v$ be a split node or a maximum in the Reeb graph $R_1$ and let $z$ be a split node or a maximum in $R_2^\eps$. Let $M^-_\eps(z) \subset R_2^\eps$ be the union of all points $x$ that are connected to $z$ by a monotone path $Q_x$ of height at most $2\eps$ and such that $f_2(q) \leq f_2(z)$, for all $q\in Q_x$. If $\phi(v) \in M^-_\eps(z)$ then $\phi^\eps(v') \in P^\eps(z)$. The analogous statement holds for a join node or a minimum. 
\end{lemma}

\begin{proof}
Set $x=\phi(v)$, then there is a monotone path $Q$ connecting $x$ to $z$ that lies below $z$.
Let $F_2^\eps$ be the function on $\mathcal{T}^{\eps}(R^\eps_2)$.
Since the height of $Q$ is at most $2\eps$, the point $(x, \eps) \in \mathcal{T}^{\eps}(R_2^\eps)$ and the point $(z,\eps-||Q||)$  (where $\eps-||Q||\geq -\eps$) are connected in the pre-image $(F_2^{\eps})^{-1}(F_2^\eps(x,\eps)) \subset \mathcal{T}^{\eps}(R_2^\eps)$, see~\autoref{fig:lemma1}.
By the definition of $\phi^\eps$, we have $\phi^\eps(v') = \pi_2^{2\eps} (\phi(v),\eps) = \pi_2^{2\eps} (x,\eps) = \pi_2^{2\eps}(z, \eps - \|Q\|)$. 
It follows that $\phi^\eps(v') \in P^\eps(z) \subset R_2^{2\eps}$.
\end{proof}

\begin{remark}
Based on \autoref{theorem:essential-labeling}, observe that the nodes with the same label move in the same direction when smoothed. We call this \textit{consistency} between the two labelings. This condition is necessary for the  distance to be a metric; see \autoref{t:exmetric} and its proof in~\autoref{sec:metric-properties-details}.
\end{remark}

\begin{theorem}
\label{t:exmetric}
Let $\mathcal{R}^L_c$ be a maximal set of isomorphism classes of consistently $L$-labeled Reeb graphs. The labeled interleaving distance is an extended metric on $\mathcal{R}^L_c$. 
\end{theorem}

We have made additional observations regarding the local neighborhood of a node $u$ under the mapping of $\eta$ {\wrt}~its $\eps$-path-neighborhood, see~\autoref{sec:path-nbhd-details}. 

\subsection{Generalizing Labeled Interleaving Between Merge Trees}\label{sec:generalized}

In this section, we show that our definition of the labeled interleaving distance between Reeb graphs generalizes the labeled interleaving distance defined in~\cite{GasparovicMunchOudot2019} for merge trees. Viewed as graphs with functions defined on them, merge trees may be considered as special cases of Reeb graphs. 
In~\cite{GasparovicMunchOudot2019}, the maximum node of a merge tree $T$ is defined to have value $\infty$ and a labeling $\lambda$ is assumed to be surjective on the set of non-root leaves. 
A matrix $\M(T_\lambda)$ called the \emph{induced matrix} is assigned to a labeled merge tree $T_ \lambda$. 
It is defined as 
\[\M(T_\lambda) [i,j] = f(\text{LCA} ( \lambda(i), \lambda(j))),\]
where $L = [n]$ and $\text{LCA}$ denotes the lowest common ancestor of nodes \edit{labeled} $i$ and $j$ in $T_\lambda$.
Given two labeled merge trees $T_{1,\lambda_1}$ and $T_{2,\lambda_2}$ with the same label set $L$, the labeled interleaving distance between the two merge trees (in the sense of~\cite{GasparovicMunchOudot2019}), denoted $\bar{d}^L_I$, is defined as 
\[ 
\bar{d}^L_I ( T_{1,\lambda_1}, T_{2,\lambda_2}) = \|\M(T_1, \lambda_1) - \M(T_2, \lambda_2) \|_{\max}.
\]
where $\|\cdot\|_{\max}$ is the element-wise maximum. 

A labeled merge tree as defined in~\cite{GasparovicMunchOudot2019} is also a labeled merge tree based on our definition (as a special case of Reeb graphs), if we replace $\infty$ by a large number depending on the function values of the tree. We assume this modification for the following proposition.

\begin{proposition}
\label{prop:generalize-mt}
For any two labeled merge trees (in the sense of~\cite{GasparovicMunchOudot2019}), we have 
\[d^L_I( T_{1,\lambda_1}, T_{2,\lambda_2})  = \bar{d}^L_I ( T_{1,\lambda_1}, T_{2,\lambda_2}).
\]
\end{proposition}

The main observation for the proof of this proposition is that for a merge tree with root at $\infty$, the smoothing operation does not change the structure of the tree, but it pushes the entire tree down uniformly.

\begin{figure}[!ht]
    \centering
    \includegraphics[width=0.6\textwidth]{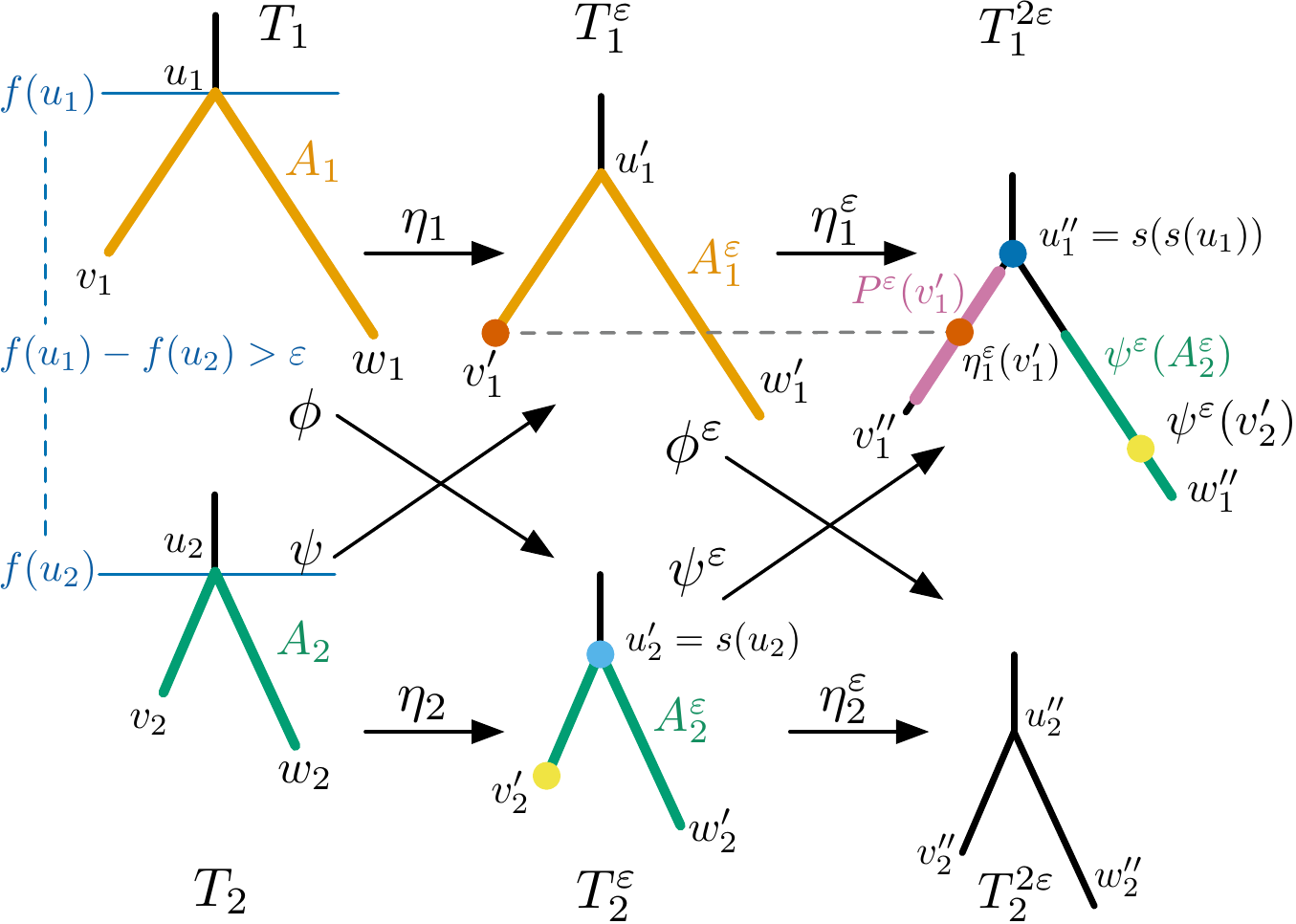}
    \caption{An illustration for the proof of~\autoref{prop:generalize-mt}.}
    \label{fig:proposition1}
\end{figure}

\begin{proof}
Since the $\infty$ nodes in the two trees have large and equal function values, the maps $\phi, \psi$ realizing $d^I_L(T_{1,\lambda_1}, T_{2,\lambda_2})$ map these nodes to the path-neighborhoods of each other. Since they do not cause any restriction, we ignore these nodes.

We first show that $d^L_I(T_{1,\lambda_1}, T_{2,\lambda_2}) \leq \bar{d}^L_I(T_{1,\lambda_1}, T_{2,\lambda_2})$. Let $\bar{\eps}=\bar{d}^L_I(T_{1,\lambda_1}, T_{2,\lambda_2})$. Since every leaf is labeled, every node is the LCA of two leaves and must appear in the matrix.
Therefore, the right hand side is the maximum of the differences between nodes with the same label, and nodes that are LCA for the same pair of labeled leaves.
We claim that there exist $\phi$, $\psi$ defining a $\bar{\eps}$-labeled interleaving between the two trees. We can define $\phi$ by mapping a leaf $v_1$ of $T_1$ to the $\bar{\eps}$-path-neighborhood of the node with the same label, say $v_2$ in $T_2$.
This is possible because $v_2$ has moved down by at least the height-difference between these nodes.
With similar reasoning, after mapping two leaves $v_1$ and $w_1$, we can also map the join node which is their LCA to the LCA of nodes with the same label in $T_2$.
The paths connecting $v_1$ and $w_1$ to their individual LCA are therefore mapped to the the smoothed tree $T_2^{\bar{\eps}}$. The map $\phi$ defined here satisfies a condition stronger than the label-preserving property ($\ref{eq:label-preserving}$), see~\autoref{remark:generalized} in~\autoref{sec:generalized}. These paths cover the whole tree, hence we have defined the morphism on all of $T_1$. Since we can consider $T_1$ as a subset of $\mathcal{T}^{\bar{\eps}}(T_2)$, and analogously $T_2$ as a subset of $\mathcal{T}^{\bar{\eps}}(T_1)$, it is easily checked that the commutativity relations~(\ref{eq:tworelations}) hold.  

Next, we show that $\bar{d}^L_I(T_{1,\lambda_1}, T_{2,\lambda_2}) \leq d^L_I(T_{1,\lambda_1}, T_{2,\lambda_2})$. Let $\eps = d^L_I(T_{1,\lambda_1}, T_{2,\lambda_2})$. 
Let $\phi$ and $\psi$ realize $d^L_I$. Let $v_1 \in T_1$ and $v_2 \in T_2$ be leaves with the same label. Since the second condition of the labeled interleaving distance is satisfied by $\phi$, we see that the height-difference between $v_1' = s(v_1)$ and $v_2' = s(v_2)$ is at most $\eps$. Since all nodes move in the same direction, the height difference between $v_1$ and $v_2$ is also at most $\eps$.

We also need to show that the height difference between unlabeled join nodes is at most $\eps$. Take two leaves $v_1, w_1 \in T_1$ and set $u_1=LCA(v_1,w_1)$. Let $A_1 \subset T$ be the union of two paths that connect $v_1$ and $w_1$ to $u_1$. Let $v_2$ and $w_2$ be the nodes in $T_2$ with the same labels as $v_1$ and $w_1$, respectively. Set $u_2= LCA(v_2,w_2)$ and let $A_2 \subset T_2$ be the union of the two paths connecting $u_2$ to $v_2$ and $w_2$.
Now observe that since we are working with merge trees, the part of $T^{\eps}_1$ other than the edge connecting to the $\infty$ node is exactly the same as $T_1$ but shifted down by $\eps$. Therefore, there exist paths $A_1^\eps \subset T_1^{\eps}$ and $A_2^\eps \subset T_2^{\eps}$ corresponding to $A_1$ and $A_2$, respectively.  

Now assume the height difference between $u_1$ and $u_2$ is larger than $\eps$. Without loss of generality, assume $u_1$ is higher than $u_2$ as in \autoref{fig:proposition1}.
Observe that now the subtree rooted at $u_2' = s(u_2)$ must be mapped by $\psi^\eps$ to a single branch going down from $u_1'' = s(s(u_1))$, and assume it is the branch containing $w_1'' = s(s(w_1))$. 
However, then $\psi^\eps(v_2')$ is also on the same branch and cannot lie on $P^\eps(v_1')$, as shown in~\autoref{fig:proposition1}. This is a contradiction.
\end{proof}

\begin{remark}\label{remark:generalized}
Observe that in the case of merge trees, we could replace the second condition of the labeled interleaving distance for Reeb graphs (\autoref{definition:labeled-interleaving}) with the following stronger condition that for all labels $\ell$ in $L$,
\begin{equation*}
    \begin{split}
    \phi(\lambda_1(\ell)) &\in P^\eps(\lambda_2(\ell)),\\ 
    \psi(\lambda_2(\ell)) &\in P^\eps(\lambda_1(\ell));
    \end{split}
\end{equation*}
where $\phi$ and $\psi$ are maps that realize an $\eps$-interleaving. 
However, we believe that with this definition, the analogous \autoref{theorem:essential-labeling} will not hold for Reeb graphs. Nevertheless, we do not make a claim at this moment and leave this question open. \autoref{lemma:strongercondition} implies that the above conditions are stronger than our label-preserving conditions~(\ref{eq:label-preserving}).
\end{remark}

\begin{remark}
The algorithm presented in this paper can be applied to the case of merge trees. However, the algorithm of \cite{GasparovicMunchOudot2019} is linear and therefore more efficient for merge trees.
\end{remark}

%% file: sec-algorithms.tex
\section{Algorithms}
\label{sec:algorithms}

\edit{We first present a general framework (\autoref{sec:framework}) for computing the labeled interleaving distance between Reeb graphs. The framework consists of two subproblems: the \emph{existence subproblem} (\autoref{sec:existence}) and the \emph{commutativity subproblem} (\autoref{sec:commutativity}).
We solve the \emph{existence subproblem} for the general case of Reeb graphs. For the \emph{commutativity subproblem}, we consider the simpler case of contour trees and give a solution that depends on the existence of unique paths between pairs of nodes. We prove that our specialized efficient algorithm for contour trees has a run-time in $\tilde{O}(n^2)$.}

\begin{figure}[!ht]
    \centering
    \includegraphics[width=0.4\linewidth]{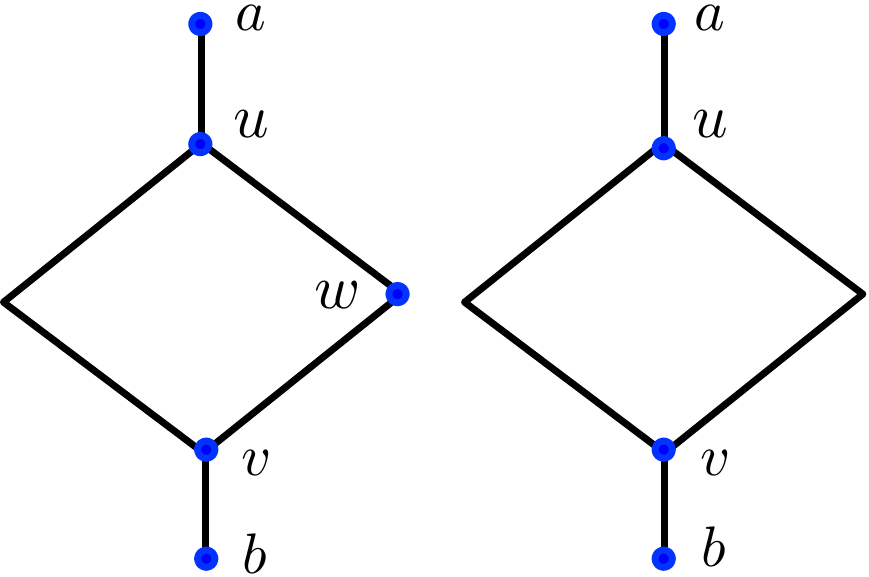}
    \caption{A spanning (left) and a non-spanning labeling (right). Labeled nodes are in blue.}
    \label{fig:spanning}
\end{figure}

When the label set $L$ is empty, the labeled interleaving distance equals the unlabeled one. Therefore, we impose a restriction on $L$ to obtain a polynomial algorithm in the case of contour trees. 
We say that a labeling $\lambda$ is \emph{spanning}, if: 
\begin{enumerate} 
\item For any pair of labeled nodes $u,v$, there is at most a single path connecting them without labeled nodes in the interior of the path;
\item All such paths cover the entire graph.
\end{enumerate} 
See \autoref{fig:spanning} for examples. By a \emph{path} we mean a simple, possibly non-monotone path in the graph. 
For instance, the labeling of a fully-labeled graph is spanning.
Furthermore, if the labeling is surjective on all leaves of a contour tree, the resulting labeling is spanning. 

If the labelings $\lambda_1$ and $\lambda_2$ are arbitrary, it is often true that the labeled interleaving distance is $\infty$.
For simplicity, we assume that the labelings are \emph{consistent}, meaning that two nodes with the same label move in the same direction when smoothed.
It is therefore not allowed that a minimum and a maximum have the same label. See~\autoref{sec:metric-properties-details} for details.

\subsection{Algorithmic Framework}
\label{sec:framework}

In order to compute the labeled interleaving distance we need to define two morphisms $\phi$ and $\psi$ satisfying the definition of labeled $\eps$-interleaving (\autoref{definition:labeled-interleaving}), such that $\eps$ is smallest with this property. 
If for some $\eps$ there is a labeled $\eps$-interleaving, then for any $\eps'>\eps$,  there is also a labeled $\eps'$-interleaving.
Our general approach is to start with a large enough $\eps$ for which a labeled interleaving exists, then to compute the infimum by a binary search over the possible values.
We explain in the following how we determine \edit{whether} a labeled $\eps$-interleaving exists for a fixed $\eps$.
We break this into two subproblems. Let $R_{1,\lambda_1}$ and $R_{2,\lambda_2}$ be the two input labeled Reeb graphs; let $R_1$ and $R_2$ be their underlying unlabeled Reeb graphs. The first subproblem is to compute for a given $\eps \geq 0$ all morphisms $\phi:R_1 \xrightarrow{} R_2^{\eps_1}$ and $\psi:R_2 \xrightarrow{} R_1^{\eps_1}$ satisfying (\ref{eq:label-preserving}). We call this subproblem the \emph{existence subproblem}. The second subproblem is to decide if the relations in (\ref{eq:tworelations}) are satisfied for maps computed in the existence subproblem. We call this the \emph{commutativity subproblem}.

The distance we are looking for depends on the function values.
We follow the convention that the complexity required to store the function values is bounded by
a fixed polynomial in the complexity of the graph. We also assume that the nodes of our Reeb graphs have distinct function values, which can be achieved by perturbation if necessary.
In this section, $n$ represents the complexity of the input, which is the total number of nodes and edges in $R_1$ and $R_2$.

\subsection{Existence Subproblem}
\label{sec:existence}

We focus on computing $\phi: R_1 \xrightarrow{} R_2^{\eps}$, and \edit{we compute} $\psi$ analogously.
Let $V_L(R_1) \subset V(R_1)$ denote the set of nodes in $R_1$ with labels, i.e., the image of $\lambda_1$. Consider $v \in V_L(R_1)$ and a label $l \in L$ such that $\lambda_1(l)=v$. Let $v'=\lambda_2(l)$. 
\edit{Then }for all labels $l \in L$, if $v=\lambda_1(l)$ and $v'=\lambda_2(l)$, $\phi^\eps$ must satisfy, 
\begin{equation}\label{eq:lid1}
    f^\eps_2(s(v')) - \eps \leq f^{2\eps}_2(\phi^\eps(s(v))) \leq f^\eps_2(s(v')) +\eps.
\end{equation}

\begin{figure}
    \centering
    \includegraphics[width=0.7\linewidth]{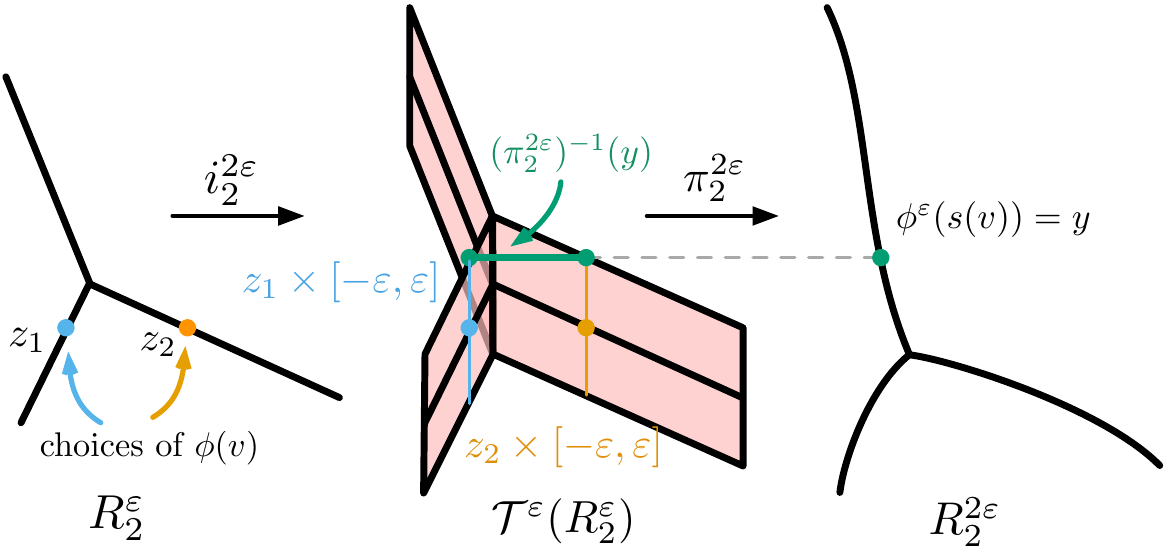}
    \caption{An illustration of the choices of $\phi(v)$.}
    \label{fig:algo-choices}
\end{figure}

Since the path-neighborhoods are strictly monotone sub-paths of the corresponding Reeb graph, if (\ref{eq:lid1}) is satisfied for $l$, $v$, and $v'$, then the point $\phi^\eps(s(v))$ is the point $x \in P^\eps_2(s(v'))$ such that $f^{2\eps}_2(x)=f^\eps_1(s(v))$. It follows that for each labeled node $v$, the image of $s(v)$ under $\phi^\eps$ is uniquely determined. 
Next, we determine all the possible points for the image of $v$ under $\phi$; there might be more than one possibility.
Recall that $\phi^\eps$ is the map induced on the Reeb graphs by a map $\Phi^\eps: \mathcal{T}^\eps (R_1) \xrightarrow{} \mathcal{T}^\eps(R_2^\eps)$ defined as $\Phi^\eps(x,t)=(\phi(x),t)$. Also recall that $s(v)$ is the projection of the point $(v,\pm \eps)$ to the Reeb graph (sign of $\eps$ depends on the type of $v$).
Let $\pi_2^{2\eps}: \mathcal{T}^{\eps}(R_2^\eps) \rightarrow R_2^{2\eps}$ and $y=\phi^\eps(s(v)) \in R_2^{2\eps}$.
The pre-image of $y$ in $\mathcal{T}^\eps(R_2^\eps)$, denoted $(\pi_2^{2\eps})^{-1}(y)$, is a collection of horizontal line segments that intersect with vertical line segments of the form $z\times [-\eps,\eps]$, where $z\in R_2^\eps$, see \autoref{fig:algo-choices}.
W.l.o.g., we assume $v$ is a split node or a maximum. From $\Phi^\eps(v,\eps)=(\phi(v),\eps)$, we know that the value of $\phi(v)$ is the midpoint of \textit{some} vertical segment that intersects the pre-image of $y$ at its highest point. In brief, the set of possible values of $\phi(v)$, denoted $C(v)$, are determined by the condition that $\phi(v)\times [-\eps, \eps]\in \mathcal{T}^\eps(R_2^\eps)$ intersects the pre-image of $y$ at $(\phi(v),\eps)$. We say that a choice of value in $C(v)$ for each $v$ result in a \textit{choice of} $\phi\!\mid_{V_L}.$
Given any $\eps$, these collections can be obtained easily in the same run time as computing a Reeb graph $R$, because we can compute a smoothing by computing the Reeb graph of $\mathcal{T}^\eps(R)$.

In the next step, we need to decide if there is a choice of images in $C(v)$ such that the resulting map can be extended to edges of $R_1$.
Since we assume that the labeling $\lambda_1$ is spanning, it follows that the graph $R_1 = Q_1 \cup Q_2 \ldots \cup Q_k$, where $Q_i$ is a simple path that starts and ends with labeled nodes $u_i$ and $w_i$. Moreover, for $i\neq j$, $Q_i$ and $Q_j$ intersect at the endpoints, if at all. 
Our goal now is to extend the morphism $\phi$ to these paths.


\begin{lemma}
\label{lemma:qi}
For a given $\eps>0$, a choice of $\phi\!\mid_{V_L}$
 can be extended to a morphism $\phi:R_1 \xrightarrow{} R^\eps_2$ if and only if for all $i$, $Q_i$ can be mapped into a path connecting $\phi^\eps(u_i)$ to $\phi^\eps(w_i)$ by a continuous, function-preserving \edit{map}.

\end{lemma}

We call the problem defined in the above lemma the \emph{path extension} problem. Apart from choices of $\phi\!\mid_{V_L}$, 
this path extension subproblem also becomes harder or easier depending on the restrictions we impose on the structure of the Reeb graphs and on the labelings. Let us denote the computational complexity of the path extension problem by $c(n)$.

\begin{proposition}
\label{p:existence}
Let $R_{1, \lambda_1}$ and $R_{2, \lambda_2}$ be labeled Reeb graphs such that the labelings are spanning. Let $p(n)$ be an upper bound on the number of choices for $\phi\!\mid_{V_L}$ for all $\eps$. There is an algorithm that computes the smallest $\eps$ for which the existence problem has a solution in time $O(n\;c(n)\;p(n)\; \log{n})$, assuming $c(n)=\Omega(\log{n})$; and in time $O(n p(n) \log^2{n})$, otherwise.
\end{proposition}

\begin{proof}
We perform a binary search on the values of $\eps$.
First we choose $\eps$ to be large enough such that the existence of the morphism $\phi$ is guaranteed.
Using (\ref{eq:lid1}), we can compute a value for $\eps$ such that the morphism $\phi$ is defined on nodes.
It is not difficult to determine here whether the distance is $\infty$.
Next we increase this $\eps$ to a value $\eps'$ such that every split node in $R_2^\eps$ is above the maximum value of $f_1$, denoted $\max(f_1)$, and every join node is below the minimum value of $f_1$, denoted $\min(f_1)$. 
For such an $\eps$, $(f_2^\eps)^{-1}([\min(f_1), \max(f_1)])$ is a monotonic arc. It is then possible to extend $\phi$ to all paths connecting labeled nodes.

Starting from $\eps'$ we execute a binary search starting from the interval $[0, \eps']$. 
During each iteration, let $[\alpha, \beta]$ be the interval at hand. We compute the midpoint $\eps = (\alpha+\beta)/2$ of the interval.
Theoretically, we can compute the Reeb graph $R_2^\eps$ using the algorithm of~\cite{Parsa2013} on the space $\mathcal{T}^\eps (R_2) = G_2 \times [-\eps, \eps]$ in $O(n \log{n})$ time, since this latter space can be turned into a simplicial complex with $O(n)$ simplices.
After computing the smoothed Reeb graph $R^\eps_2$ we use the path extension subroutine and spend $O(np(n))$ time to check if each choice of the morphism $\phi$ defined on labeled nodes can be extended to all paths connecting labeled nodes. If the extension is possible for all paths for some choice, we set $\beta = \eps$; otherwise, we set $\alpha = \eps$. This step takes $O(nc(n)p(n)+ n\log n)$ time to execute.

To finish the proof, we need to argue that the binary search stops after $O(\log{n})$ iterations. Observe that as we smooth the Reeb graphs, the merge and split nodes of both input graphs move down and up, respectively.
Therefore, we need to take careful consideration when the function values of the two sets of nodes cross each other.
Then the value of $\eps$ we are looking for has the same asymptotic complexity as the input function values, which is at most $O(n)$ by our assumptions. Hence, we need at most $O(\log{n})$ steps in the binary search.
\end{proof}

\subsection{Commutativity Subproblem}
\label{sec:commutativity}
In \autoref{sec:existence}, we have computed $\eps$ such that certain morphisms $\phi$ and $ \psi$ exist. However, there might not be a unique way of mapping a path and the choice might be important for satisfying the commutativity requirement. If the first Betti numbers $\beta_1(G_i)$, for $i=1,2$, is $O(\log{n})$, then there are polynomially many paths to choose from. 
We consider a simpler setting when the Reeb graphs are trees, i.e., contour trees. In this case, there is a unique path connecting any two vertices. Using this property we can prove that for a given $\eps$ there is at most a single label-preserving morphism. We then check this morphism if it satisfies the commutativity relations. We thus have the following theorem.

\begin{theorem}
\label{theorem:compute}
Let $R_{1, \lambda_1}$ and $R_{2, \lambda_2}$ be contour trees. Assume $\lambda_1, \lambda_2$ are consistent labelings that cover the leaves of the contour trees. Then there is an algorithm for computing the labeled interleaving distance in time $O(n^2\log^2{n})$.  
\end{theorem}

\begin{proof}
First, we claim that if the Reeb graph is a tree then $p(n)=1$, and thus there is at most one choice for $\phi\!\mid_{V_L}$. Let $w$ be the global maximum of $R_2$. Since $w$ is a labeled node, some node $v \in R_1$ has the same label as $w$. We know the point $\phi^\eps(s(v))$ and the point $\phi(v)$ is one of the points in $C(v)$. These choices are caused by the existence of a join node below $s(w)$ as shown in \autoref{fig:algo-choices}, where $s(w)$ could be the maximum of $R_2^\eps$.
Since a join node moves down during smoothing, it can only have moved down from above $w$, which is not possible. Hence, there must be a single choice in $C(v)$. We thus defined $\phi$ on a single node uniquely. We show now that this morphism extends uniquely to all of $R_1$, if at all.

We define $\phi$ on the sub-tree $T$ rooted at $v$ that consists only of join nodes. We move down from $v$ inductively, and at each step, we add a monotone path $P$ that has either only split nodes in the interior, or no node at all.
Assume that we have already extended our map to some sub-tree of $T$. Let $P$ connect nodes $v_1$ and $v_2$ such that the morphism is uniquely defined on $v_1$. We want to extend the map uniquely to $P$ or deduce that such an extension is not possible.
There could be multiple choices for $v_2$ in $C(v_2)$. Since $v_2$ moves down when smoothed, these choices must be caused by some split node in $R^\eps_2$. For the sake of contradiction, suppose there are two points $q_1, q_2 \in C(v_2)$ such that there is a monotone path from $\phi(v_1)$ to $q_i$, $i=1,2$. These paths have higher function values than $C(v_1)$, i.e, $\phi(v_1)$.
There is also a split node that connects $q_1$ and $q_2$ and has function value less than $C(v_2)$. This is however impossible, since $R^\eps_2$ is a tree. Therefore, there is at most one point in $C(v_2)$ that we can connect to. It follows that we can extend to $T$ uniquely, if at all.

We now have nodes in $T$ with $\phi$ defined on them. We can now start moving up from these nodes and define the function using a similar argument as above. These are split nodes in the interior of the path $P$ and can possibly include $v$ itself.
We continue extending the function $\phi$ as long as there are split and join sub-trees remaining whose roots have been already uniquely mapped. This completes the proof of our claim that $p(n)=1$.

Since the labelings are spanning, we can solve the existence subproblem as in~\autoref{p:existence}. To solve the path extension subproblem, we only need to check the possibility of mapping a path to another unique path, which takes $O(n)$ time per edge.
As there is a unique way of mapping any edge, we need to perform a binary search as in~\autoref{p:existence} to find the smallest $\eps$ such that the commutativity relations hold for the same maps found from existence subproblem. We can do this simultaneously as we solve the existence problem.
Checking the commutativity can be done in linear time per edge, and therefore in time $O(n^2)$ for every step of the binary search. The entire algorithm takes $O(n^2 \log{n}) = \tilde{O}(n^2)$ time.
\end{proof}

%% file: sec-geodesic.tex
\section{Nonexistence of Geodesics}
\label{sec:counterexamples}

As explained in~\autoref{sec:introduction}, it is important to know whether a given metric on Reeb graphs (or contour trees) is intrinsic, i.e.,~whether the metric space has geodesics. 
In~\cite{DeSilvaMunchPatel2016}, the labeled interleaving distance between merge trees is used to argue that the (ordinary) interleaving distance between merge trees is strictly intrinsic. Therefore, the space of merge trees under this metric is intrinsic.
\edit{Here, we continue our discussion in \autoref{sec:generalized} on generalizing labeled interleaving between merge trees to Reeb graphs. Specifically, we consider this intrinsic property for the interleaving distance for Reeb graphs and contour trees.}
We show that the (ordinary) interleaving distance is not strictly intrinsic in these more general settings. \edit{Therefore, the labeled interleaving distance must not be strictly intrinsic as well.}
Our counterexamples are simple, however, the fact that they are counterexamples is not immediate. 

If a space is intrinsic, there is an ``average'' for any two objects in the space. 
That is, for all pairs of Reeb graphs $R_1$ and $R_2$, there exists an average Reeb graph $M=M(R_1,R_2)$ such that $d_I(M,R_1) = d_I(M,R_2)= d_I(R_1,R_2)/2$. We show below that the statement is not true for Reeb graphs or contour trees. 

\subsection{Counterexample for Reeb Graphs}
\label{sec:countereexample-rg}

\begin{figure}[!h]
    \centering
    \includegraphics[width=0.7\columnwidth]{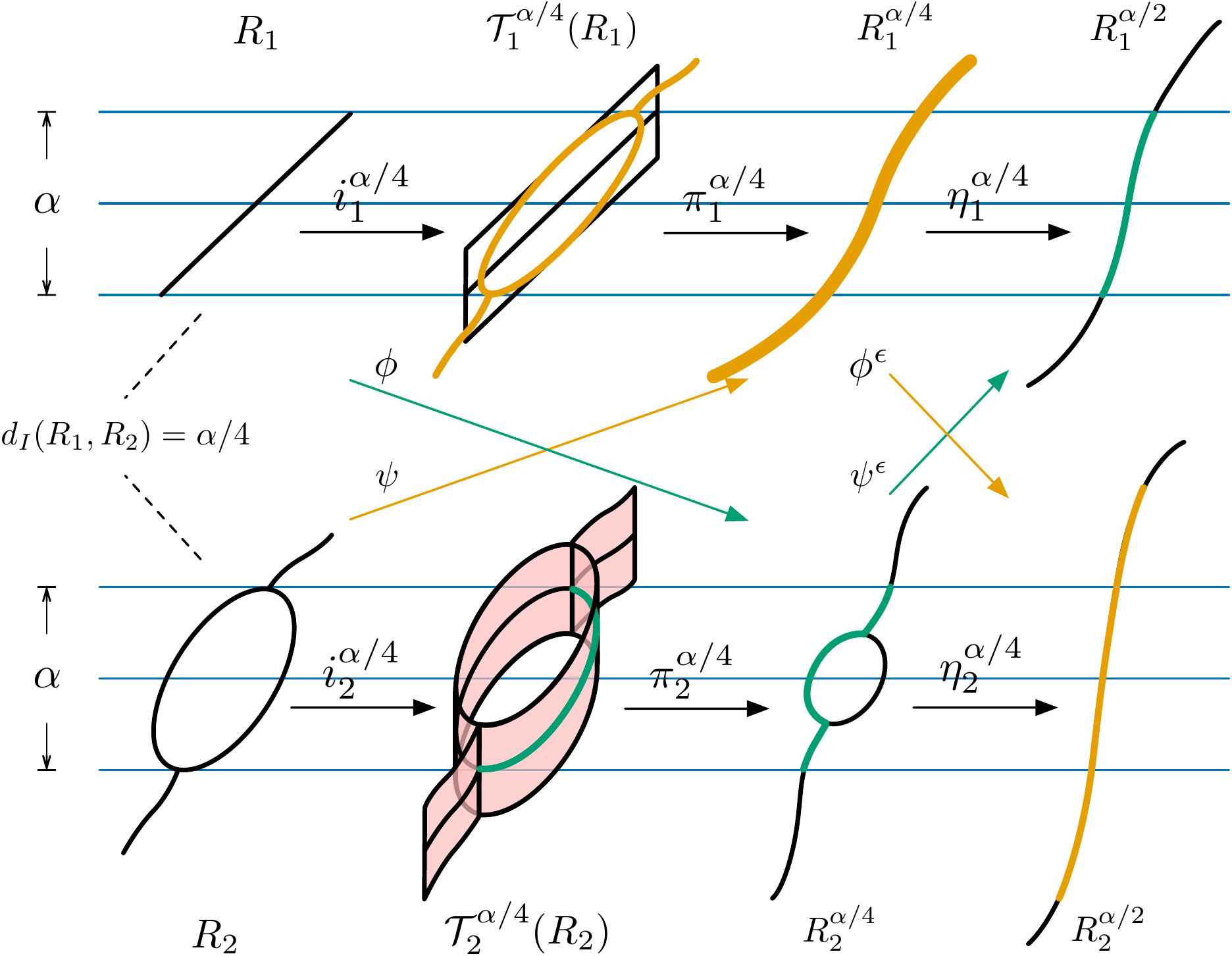}
    \caption{A counterexample for Reeb graphs.}
    \label{fig:counterReeb}
\end{figure}

Let $R_1$ be a line segment of height $\alpha$ and let $R_2$ be a simple loop of height $\alpha$. The morphisms drawn in \autoref{fig:counterReeb} show that $d_I(R_1, R_2) = \alpha/4$, because of the loop of height $\alpha$ in $R_2$.
 
We assume for the sake of contradiction that there is a Reeb graph $M$ such that $d_I(M,R_1)=d_I(M,R_2)=\alpha/8 = \eps$.
We have the following commutative diagram, 

\begin{equation*}
  \begin{tikzcd}
  M
    \ar[r]
    \ar[dr, "{}", very near start, outer sep = -2pt]
  & M^{\alpha/8}
    \ar[r]
    \ar[dr, "{}", very near start, outer sep = -2pt]
  & M^{\alpha/4}
  \\
  R_2
    \ar[r]
    \ar[ur, crossing over, "{}"', very near start, outer sep = -2pt]
  & R_2^{\alpha/8}
    \ar[r]
    \ar[ur, crossing over, "{}"', very near start, outer sep = -2pt]
  & R_2^{\alpha/4}
  \end{tikzcd}
\end{equation*}

Since $R_2$ and $R_2^{\alpha/4}$ each contains a loop of height $\alpha$ and $\alpha/2$, respectively, and the above diagram is commutative, $M^{\alpha/8}$ must contain a loop of height at least $\alpha/2$.
Then by \autoref{l:loopcontract}, $M$ must have a loop of height at least $\alpha/2+\alpha/4 = 3\alpha/4$. Now consider the relationship between $R_1$ and $M$. Since $R_1$ is a line segment and $M$ has a loop of height at least $3\alpha/4$, the interleaving distance $d_I(M, R_1)$ is at least $(3\alpha/4)/4 = 3\alpha/16$.
This is because that the loop in $M$ must disappear after smoothing by $2\eps$, similar to the example in \autoref{fig:counterReeb}.
However, $3\alpha/16 > \alpha/8 = \eps$, contradicting our assumption that $d_I(M, R_1) = d_I(M, R_2) = \alpha/8$.

\begin{lemma}
\label{l:loopcontract}
Let $R$ be a Reeb graph. If $R^\eps$ has a loop of height $a$, then $R$ must contain a loop of height $a+2\eps$.
\end{lemma}

\begin{proof}
Let $r$ be an injective, function-preserving map $r: R^\eps \to \mathcal{T}^\eps(R)$. $r$ exists by our assumption that nodes have distinct function values.
Let $O$ be the loop of $R^\eps$ with height $a$, and let $x \in R^\eps$ be a point of $O$. Let $y = r(x)$ and $J = z\times [-\eps, \eps] \subset \mathcal{T}^\eps(R)$ be the vertical interval containing $y$. We move $y$ up or down until it coincides with $z\times\{0\} \in R \times \{0\}$. We perform this move for all $x \in O$ such that $x$ is a vertex of $R^\eps$ or $z$ is a vertex of $R$.   

The highest point of $O$, $\max(O)$, is a join node and its image must move $\eps$ up to lie in $R\times\{0\}$.
Similarly, the lowest point of $O$, $\min(O)$, is a split node and its image must move down $\eps$ to lie in $R \times\{0\}$. Other nodes move at most $\eps$.
Now, $\max(O)$ and $\min(O)$ are the highest and lowest nodes of a new loop with height $a + 2\eps$.
\end{proof}


\subsection{Counterexample for Contour Trees}
\label{sec:countereexample-ct}

Since the loop in \autoref{sec:countereexample-rg} is essential, one might conjecture that when there is no loop in a Reeb graph (i.e.,~for a contour tree), the geodesic property holds.
In this section, we present a counterexample to this conjecture.

\begin{figure}[!h]
    \centering
    \includegraphics[width=0.7\textwidth]{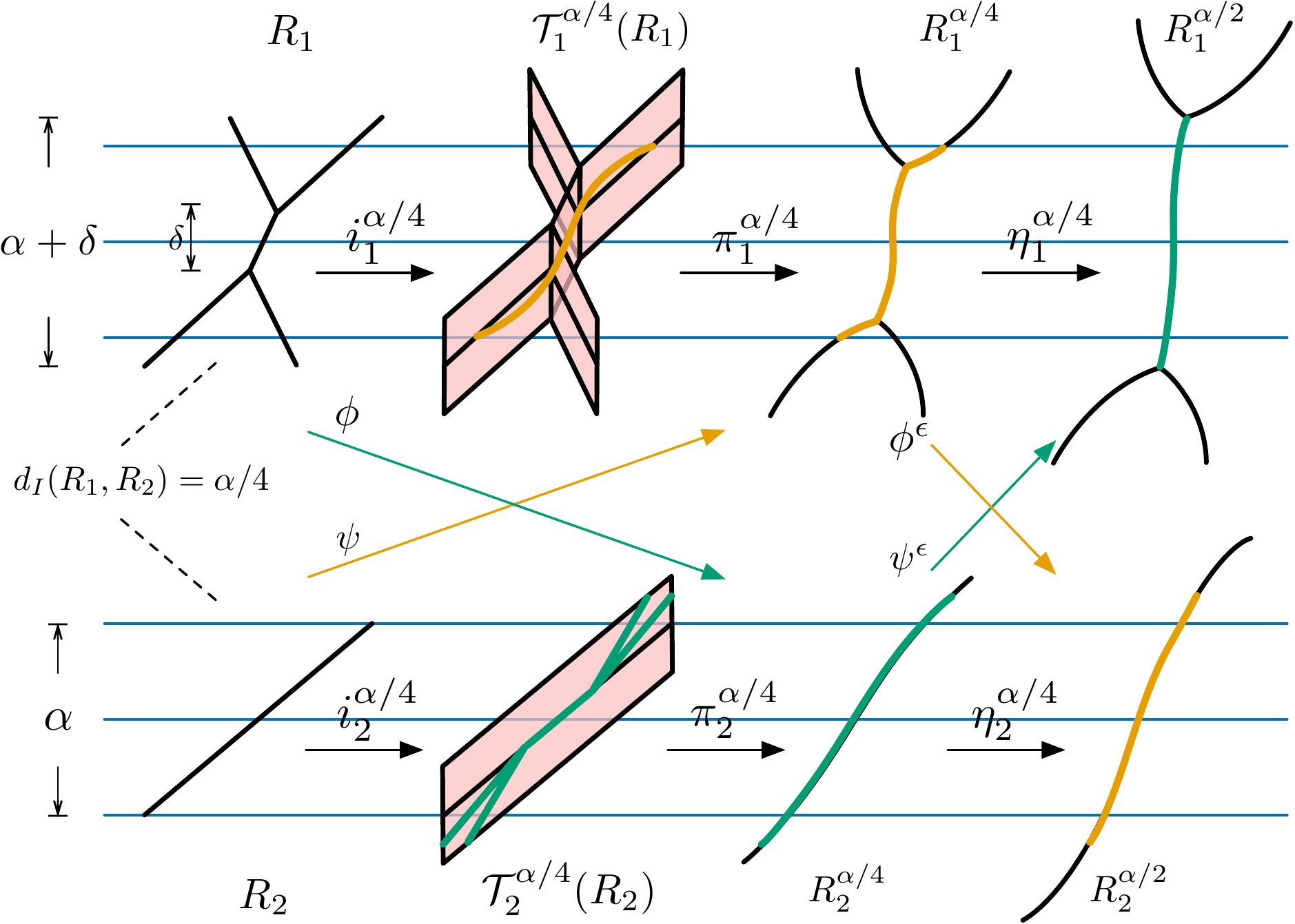}
    \caption{A counterexample for contour trees.}
    \label{fig:counterContour}
\end{figure}

Let $R_1$ be an X-shaped contour tree in \autoref{fig:counterContour} (top), where we consider the crossing point to be the superposition of a join node and a split node that are arbitrarily close.  
Let $R_2$ be a line segment in \autoref{fig:counterContour} (bottom). 
The total height of the Reeb graphs are equal and we denote it by $\alpha$. Let the lowest point of
these graphs be the origin of the height function and let the line segment be centered at $\alpha/2$.
Observe that $d_I(R_1,R_2)=\alpha/4$. The maps $\phi:R_1 \to R_2^\eps$ and $\psi: R_2 \to R_1^\eps$ defining an $\alpha/4$-interleaving are drawn in \autoref{fig:counterContour}. 
We again consider the following commutative diagram:

\begin{equation*}
  \begin{tikzcd}
  M
    \ar[r]
    \ar[dr, "{}", very near start, outer sep = -2pt]
  & M^{\alpha/8}
    \ar[r]
    \ar[dr, "{}", very near start, outer sep = -2pt]
  & M^{\alpha/4}
  \\
  R_1
    \ar[r]
    \ar[ur, crossing over, "{}"', very near start, outer sep = -2pt]
  & R_1^{\alpha/8}
    \ar[r]
    \ar[ur, crossing over, "{}"', very near start, outer sep = -2pt]
  & R_1^{\alpha/4}
  \end{tikzcd}
\end{equation*}

\begin{figure}[!ht]
    \centering
    \includegraphics[width=0.8\linewidth]{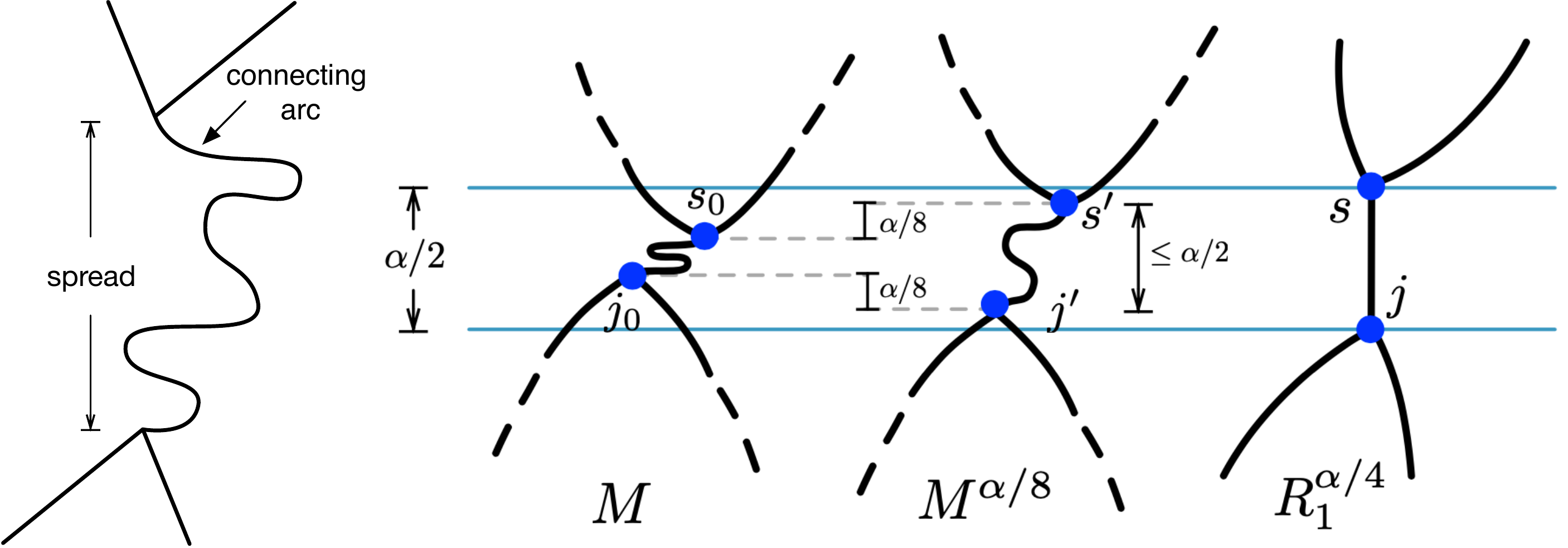}
    \caption{Left: A joint-split structure. Right: additional illustration for the counterexample involving contour trees.}
    \label{fig:counterContour-M}
\end{figure}

For the sake of contradiction, we assume there is a contour tree $M$ such that $d_I(M,R_1)=d_I(M,R_2)=\alpha/8$.
We call a (part of a) contour tree a \emph{join-split structure}, or JS-structure \footnote{This term is not intended to be related to the one in~\cite{WangWangWenger2014}.}, if it consists of a split and a join node connected by a \emph{connecting arc}, see \autoref{fig:counterContour-M} (left).
We call the function (height) value difference between the split and the join node the \emph{spread} of the JS-structure. Note that the spread can be negative. For instance, $R_1$ in \autoref{fig:counterContour} is a JS-structure where the connecting arc is trivial and the spread is $0$.

As illustrated in~\autoref{fig:counterContour-M} (right), the tree $R_1^{\alpha/4}$ is a JS-structure with spread $\alpha/2$. Let $j$ and $s$ denote the join and split nodes in $R_1^{\alpha/4}$.
Since the above diagram commutes, there must be a split node $s'$ in $M^{\alpha/8}$ lying at or below $s$ and a join node $j'$ lying at or above $j$. Then, there is a $JS$-structure in $M^{\alpha/8}$ of spread at most $\alpha/2$.
Since there are corresponding join and split nodes $j_0, s_0 \in M$ with function values $\alpha/8$ above $j'$ and below $s'$, respectively, there must be a $JS$-structure in $M$ of spread at most $\alpha/4$.
Since $R_2$ is a line segment, if $d_I(M, R_2) = \alpha/8 = \eps$, then after moving $2\eps$ down, the join node $j_0$ must have a function value of at most $0$; and after moving $2\eps$ up, the split node $s_0$ must have a function value of at least $\alpha$.
Then $s_0$ must lie at most $\alpha/4$ below $\alpha$ and $j_0$ must lie at most $\alpha/4$ above $0$. It follows that the spread is at least $\alpha/2$, which contradicts our earlier assumption.

\begin{remark}
In the above example, the join and split nodes in $R_1$ coincide. However, this is not essential and they can be separated by $\delta$ where $\delta$ is sufficiently small but positive. For example, it may be a Reeb graph of a Morse function defined on a 2-manifold.
\end{remark}

\subsection{More on Metric Properties of the Space of Reeb Graphs}
The counterexamples above show that the spaces of Reeb graphs and contour trees are not geodesic \wrt~$d_I$, implying that $d_I$ is not strictly intrinsic. We can also deduce the following,

\begin{proposition}
\label{prop:not-intrinsic}
The interleaving distance $d_I$ is not intrinsic.
\end{proposition}

\begin{proof}
If the interleaving distance was intrinsic, then there would exist paths connecting the two Reeb graphs $R_1$ and $R_2$ whose lengths, computed using the metric $d_I$, approach the interleaving distance arbitrarily closely. Then there would exist Reeb graphs which are arbitrarily close to a midpoint of $R_1$ and $R_2$ \wrt~$d_I$.
However, this cannot be true by the above example.
\end{proof}

In the examples presented above, the parameter $\alpha$ is arbitrary. It follows that there are counterexamples that are arbitrarily close to each other \wrt~$d_I$.
In \cite{DeSilvaMunchPatel2016}, the authors asked if $\hat{d}_I$ (and other intrinsic versions of distances on Reeb graphs) is strictly intrinsic. We conjecture that $\hat{d}_I$ is not strictly intrinsic, and more generally we conjecture the following,

\begin{conjecture}
\label{conj}
The space of isomorphism classes of Reeb graphs with any metric $d$ is not a geodesic space.
\end{conjecture}

%% file: sec-conclusion.tex
\section{Conclusion and Discussion}
\label{sec:discussion}

In this paper, we define a labeled interleaving distance for Reeb graphs and prove its properties. 
The labeled interleaving distance does not approximate the ordinary interleaving distance, nor does it help with its computation. 
Rather, the labeled interleaving distance may be a substitute for the ordinary interleaving distance in some applications. 
We also show that the ordinary interleaving distance between Reeb graphs is not intrinsic. 
Therefore, there is no interpolation or average Reeb graph {\wrt} this distance between two close enough Reeb graphs. 
There are at least two ways to tackle this issue. First, we could try to find a Reeb graph which approximates an average. This would require a careful determination of the criteria for this average, and whether we should give more importance to one Reeb graph compared to the other one. 
Second, we could try to come up with distances with intrinsic properties. 
Even if \autoref{conj} is true, there might be useful pseudo-metrics.

We present a first algorithm for computing the labeled interleaving distance between contour trees in polynomial time.  We do not know whether a more efficient algorithm exists. 
It would also be useful to know if a polynomial algorithm exists for labeled Reeb graphs. 
Chambers \etal~\cite{ChambersMunchOphelders2021} recently defined a family of metrics related to the interleaving distance on Reeb graphs, using the concept of smoothing and truncation. 
Our current labeled interleaving distance only uses the concept of smoothing. 
Understanding the intrinsic properties of these truncated interleaving distances could also be an interesting continuation of the present work.

%% file: sec-interleaving-details.tex
\section{An Observation on Path Neighborhoods}
\label{sec:path-nbhd-details}

The following is an observation which we do not use but it might find application elsewhere.

\begin{lemma}
\label{lemma:Neps}
Let $u$ be a node in a Reeb graph $R$ and let $N_\eps(u)$ be the union of all points that are connected to $u$ by a path $Q$ of height at most $\eps$ and such that $f(q) \geq f(u)$ for all $q \in Q$, or $f(q) \leq f(u)$ for all $q\in Q$. Then $\eta(N_\eps(u)) \subset P^\eps(u)$. 
\end{lemma}

\begin{figure}[!h]
    \centering
    \includegraphics[width=0.7\textwidth]{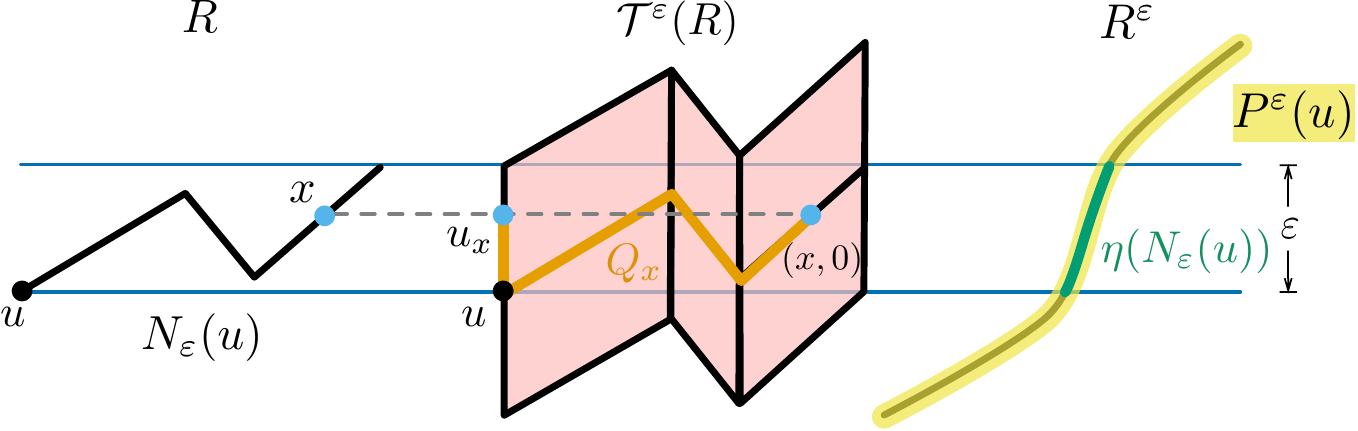}
    \caption{An illustration for the proof of~\autoref{lemma:Neps}.}
    \label{fig:lemma2}
\end{figure}

\begin{proof} 
Take $x \in N_\eps(u)$. By assumption, the point $x$ is connected by some path $Q_x \subset Q$ to $u$ such that $Q_x$ lies completely above or below $u$ in the function values.
This is because for a path to attain the value $f(u)$, it must be crossing $f(u)$ or be incident to $u$.
Consider the thickening $\mathcal{T}^\eps(R)$ and the function $f^\eps: \mathcal{T}^{\eps}(R) \xrightarrow[]{} \mathbb{R}$.
Since $||Q_x||\leq \eps$, there is a point $u_x \in u \times [-\eps, \eps]$ such that $f^\eps(u_x)=f^\eps((x,0))=f(x)$, see~\autoref{fig:lemma2}. 
We claim that $x$ and $u_x$ are connected in the pre-image $(f^{\eps})^{-1}(f(x))$. Consider the strip $Q_x \times [-\eps, \eps]$. The range $[f(u), f(u_x)]$ is always covered by lines $q\times [-\eps, \eps]$, for $q\in Q_x$. 
\end{proof}

\section{Metric Properties of the Labeled Interleaving Distance}
\label{sec:metric-properties-details}

\begin{figure}[!h]
    \centering
    \includegraphics[width=0.1\columnwidth]{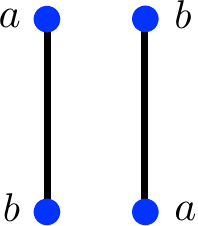}
    \caption{A pair of inconsistently labeled Reeb graphs}
    \label{fig:infinite-dist}
\end{figure}

It is easy to see that the labeled interleaving distance is not necessarily finite.
This happens when the nodes that are labeled by the same label are originally apart and also move in opposite directions when smoothed, see~\autoref{fig:infinite-dist} for an example. Consequently, it could potentially only be an extended metric on the isomorphism class of labeled Reeb graphs, with the same label set $L$. 

In this section, we first put a condition on possible labelings and show that under this condition the labeled interleaving distance becomes a metric. Since the label set $L$ is arbitrary, this provides a large number of metrics on large subsets of Reeb graphs that could be of interest in many applications, both in theory and in practice.

For a label set $L$, we denote the set of all $L$-labeled Reeb graphs by $\mathcal{R}^L$. We say that two labelled Reeb graphs $R_{1,\lambda_1}$, $R_{2,\lambda_2}$ are \textit{consistently labeled } if for each label $\ell \in L$, the nodes $v_1=\lambda_1(\ell)$ and $v_2=\lambda_2(\ell)$ move in the same direction when smoothed. In other words, if $v_1$ is a maximum or split, then $v_2$ is also a maximum or a split, and if $v_1$ is a minimum or a join, then $v_2$ is a minimum or a join. A set of labeled Reeb graph $\mathcal{R} \subset \mathcal{R}^L$ is called \textit{consistently labeled} if each pair of labeled Reeb graphs in it are consistently labeled.   


The following lemma will be used in the proof of \autoref{t:exmetric}.

\begin{lemma}\label{lemma:strongercondition}
\label{l:lemma-const-m}
Let $R_1$ and $R_2$ be two Reeb graphs, and $v_1 \in R_1$, $v_2 \in R_2$ be two nodes such that they move in the same direction when smoothed. Let $\phi: R_1 \xrightarrow{} R_2^\epsilon$ be a morphism satisfying
$\phi(v_1) \in P^\eps (v_2).$ Then, for any $\delta \geq 0$,
$$ \phi^\delta(s_{\delta}(v_1)) \in P^\eps (s_{\delta}(v_2)),$$
where $s_\delta$ denotes the correspondence $s_\delta: V(R) \rightarrow V(R^\delta)$ for any pair of Reeb graphs.
\end{lemma}

\begin{figure}[!h]
    \centering
    \includegraphics[width=0.9\columnwidth]{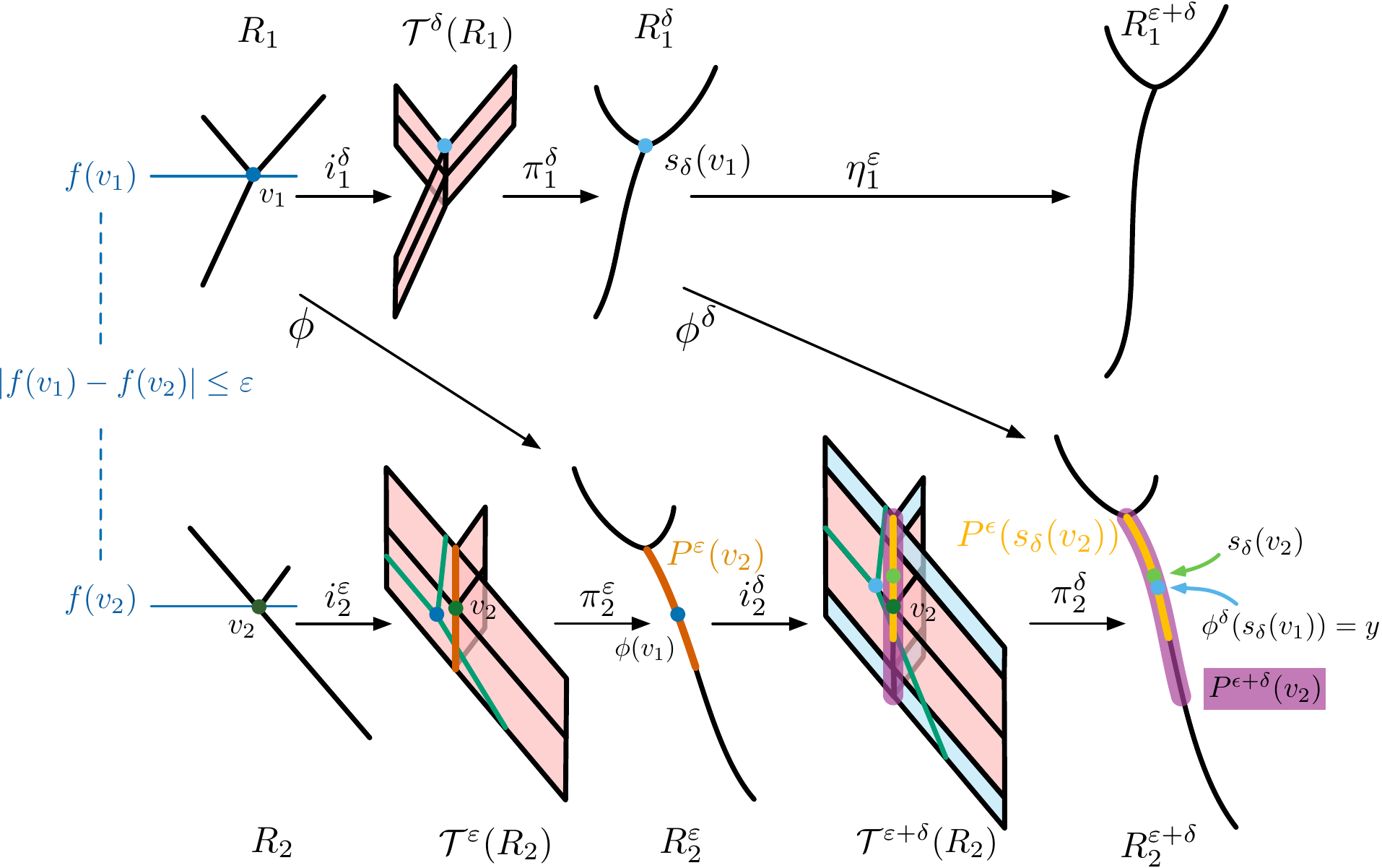}
    \caption{An illustration for the proof of \autoref{l:lemma-const-m}.}
    \label{fig:lemmaconstm}
\end{figure}

\begin{proof}
Without loss of generality, assume that both $v_1$ and $v_2$ are maxima or split nodes. The condition  $\phi(v_1) \in P^\eps (v_2)$ implies that $\vert f_1(v_1)-f_2(v_2) \vert \leq \eps$. Since the nodes move in the same direction we have 
$\lvert f_1^\delta(s_{\delta}(v_1)) - f_2^\delta(s_{\delta}(v_2)) \rvert \leq \eps$.

On the other hand, we know that $y=\phi^\delta(s(v_1)) \in P^{\epsilon+\delta}(v_2)$, and $y$ is at the function distance at most $\eps$ from $s_{\delta}(v_2)$. 
Hence, $y$ has to be in $P^\eps(s_{\delta}(v_2))$, since this path neighborhood is the upper part of $P^{\eps+\delta}(v_2)$. See \autoref{fig:lemmaconstm} for an example.  
\end{proof}

\begin{proof}[Proof of~\autoref{t:exmetric}]
All conditions of $R_c^L$ being a metric are trivial to check except for the triangle inequality, for which we provide a proof. Let $R_1, R_2, R_3 \in \mathcal{R}^L_c$ be Reeb graphs such that $d^L_I(R_1, R_2) = \eps_1$ and $d^L_I(R_2, R_3)=\eps_2$. We need to prove that $d^L_I(R_1, R_3)\leq \eps_1+\eps_2$.
For this purpose, it is enough to show that there is a labeled $\eps_1+\eps_2$-interleaving between $R_1$ and $R_3$. Let $\phi_1, \psi_1$ define a labeled $\eps$-interleaving between $R_1$ and $R_2$, and $\phi_2, \psi_2$ be analogous for $R_2$ and $R_3$. We need to define $\phi: R_1 \xrightarrow{} R_3^{\eps_1+\eps_2}$, and $\psi: R_3 \xrightarrow{} R_1^{\eps_1+\eps_2}$.
We take $\phi = \phi_2^{\eps_1} \circ \phi_1$ and $\psi = \psi_1^{\eps_2} \circ \psi_2$ to be the compositions. As in \cite{DeSilvaMunchPatel2016}, Proposition 4.5, this pair of morphisms define an $\eps_1+\eps_2$-interleaving.
Here we show that these morphisms satisfy the additional labeling requirement, namely, $$\phi^{\eps_1+\eps_2}(s(v_1)) \in P^{\eps_1+\eps_2}(s(v_3))$$ when $v_1$ and $v_3$ have the same label, and analogously for $\psi$. We focus on $\phi$. 

Let $v_1 \in R_1$, $v_2 \in R_2$ and $v_3 \in R_3$ be nodes with the same label. First, note that 
$$\phi^{\eps_1+\eps_2} = (\phi_2^{\eps_1}\phi_1)^{\eps_1+\eps_2} = \phi_2^{2\eps_1+\eps_2} \circ \phi_1^{\eps_1+\eps_2}.$$
    
Second, using~\autoref{l:lemma-const-m} with $\delta = \eps_2$, 
$$\phi_1^{\eps_1}(s_{\eps_1} (v_1)) \in P^{\eps_1}(s_{\eps_1} (v_2))$$ implies $$y=\phi_1^{\eps_1+\eps_2}(s_{\eps_1+\eps_2} (v_1)) \in P^{\eps_1} (s_{\eps_1+\eps_2}(v_2)).$$ 

Similarly, $\phi_2^{\eps_2}(s_{\eps_2} (v_2)) \in P^{\eps_2}(s_{\eps_2} (v_3))$ 
implies $$\phi_2^{\eps_1+\eps_2}(s_{\eps_1+\eps_2} (v_2)) \in P^{\eps_2} (s_{\eps_1+\eps_2}(v_3)).$$ 
By "thickening" the two sides of the above we obtain, 
$$\phi_2^{2\eps_1+\eps_2}(P^{\eps_1} (s_{\eps_1+\eps_2} (v_2)) ) \subset P^{\eps_1+\eps_2}(s_{\eps_1+\eps_2}(v_3)).$$ 

Putting these together, 
$$\phi^{\eps_1+\eps_2}(s_{\eps_1+\eps_2}(v_1)) = \phi_2^{2\eps_1+\eps_2}(y) \in  \phi_2^{2\eps_1+\eps_2}(P^\eps_1 (s_{\eps_1+\eps_2} (v_2))) \subset P^{\eps_1+\eps_2}(s_{\eps_1+\eps_2}(v_3)),$$
which is what we wanted to prove.  
\end{proof}